\documentstyle[12pt]{article}

\def\vcb{\left| V_{cb} \right|}
\def\vtd{\left| V_{td} \right|}
\def\vub{\left| V_{ub}/V_{cb} \right|}

\newcommand{\Bsg}{$B \to X_s \gamma$ }

\newcommand{\bea}{\begin{eqnarray}}
\newcommand{\eea}{\end{eqnarray}}
\newcommand{\bd}{\begin{displaymath}}
\newcommand{\ed}{\end{displaymath}}

\begin{document}
\thispagestyle{empty}
\begin{flushright}
 MPI-PhT/95-88 \\
 TUM-T31-97/95 \\
hep-ph/9509329 \\
 September 1995
\end{flushright}
\vskip1truecm
\centerline{\Large\bf  Theoretical Review of B-Physics
   \footnote[1]{\noindent Based on two invited talks given at
    "Beauty 95", Oxford, July 9--14, 1995, to appear in the proceedings.\\
   Supported by the German
   Bundesministerium f\"ur Bildung und Forschung under contract
   06 TM 743 and by the CEC science project SC1--CT91--0729.\\
    e-mail: buras @ feynman.t30.physik.tu-muenchen.de}}
\vskip1truecm
\centerline{\sc Andrzej J. Buras}
\bigskip
\centerline{\sl Technische Universit\"at M\"unchen, Physik Department}
\centerline{\sl D-85748 Garching, Germany}
\vskip0.6truecm
\centerline{\sl Max-Planck-Institut f\"ur Physik}
\centerline{\sl  -- Werner-Heisenberg-Institut --}
\centerline{\sl F\"ohringer Ring 6, D-80805 M\"unchen, Germany}

\vskip1truecm
\centerline{\bf Abstract}

We review several aspects of B-Physics. In particular we discuss:
i) The theoretical framework for B-decays, ii) Weak decays beyond
leading logarithms, iii) Standard analysis of the unitarity triangle,
iv) Rare B-decays, v) CP violation including the issue of electroweak
penguins and  vi) Future visions of this field.

\newpage

\setcounter{page}{1}

\centerline{\Large\bf  Theoretical Review of B-Physics}
 \vskip1truecm
\centerline{\sc Andrzej J. Buras}
\bigskip
\centerline{\sl Technische Universit\"at M\"unchen, Physik Department}
\centerline{\sl D-85748 Garching, Germany}
\vskip0.6truecm
\centerline{\sl Max-Planck-Institut f\"ur Physik}
\centerline{\sl  -- Werner-Heisenberg-Institut --}
\centerline{\sl F\"ohringer Ring 6, D-80805 M\"unchen, Germany}
\vskip1truecm
\centerline{\bf Abstract}
{\small
We review several aspects of B-Physics. In particular we discuss:
i) The theoretical framework for B-decays, ii) Weak decays beyond
leading logarithms, iii) Standard analysis of the unitarity triangle,
iv) Rare B-decays, v) CP violation including the issue of electroweak
penguins and  vi) Future visions of this field.
}

\section{Introduction}

The theoretical review of B-physics presented below is based on two talks
I have given at ``Beauty 95'' held in Oxford in July this year. The first talk
dealt with the existing theoretical framework for B-decays. The second
talk, the final talk of the workshop, was a theoretical review of B-physics.

It is
not my intention to discuss all aspects of B-physics here.
I would rather like
to concentrate on certain topics in B-physics which I find to be
particularly interesting and which I expect to play an important role
during this and the following decade. Undoubtly the choice of these
topics is dictated to a large extent by my own research in this field,
but afterall this is the part of B-physics which I know best and
consequently there is some probability that certain messages given
below will turn out to be useful. I have organized the material
as follows:

Section 2 deals with the theoretical framework for B-decays which
is based on the operator product expansion (OPE), the renormalization
group, Heavy Quark Effective Theory (HQET) and Heavy Quark Expansions (HQE).
Only main ideas are presented there and the technical details
are left out. A brief description of the "penguin-box expansion" (PBE)
is also given. The latter can be regarded as a version of the OPE which
is particularly useful for the study of the $m_t$ dependence in weak decays.
Finally the status of higher order QCD corrections in weak decays is
briefly summarized.

Section 3 discusses the by now standard analysis of the unitarity triangle
(UT). After recalling the Wolfenstein parametrization and several basic
formulae, I will collect a few messages which should be
useful for the UT practitioners. Next the input parameters:
$\mid V_{cb} \mid $,
$\mid V_{ub}/V_{cb}\mid $, $B_K$, ${F_B\sqrt{B_B}}$ and $m_t$ will be
discussed.
Finally results for the UT and related quantities of interest will
be presented.

Section 4 deals with three topics: the semi-leptonic branching ratio
($B_{SL}$), the number of charmed hadrons per B-decay ($n_c$) and
the question of factorization in non-leptonic B-decays.

Section 5 summarizes the present status of the three stars in the field of
rare B decays: $B \to X_s\gamma$, $B \to X_s e^+e^-$ and $B \to \mu\bar\mu$.

Section 6 can be regarded as an express review of CP violation in B-decays.
After a list of decays which allow theoretically clean determinations of
CKM phases, I will in particular discuss the hot subject of this year:
the issue of electroweak penguins in B-decays.

Section 7 is an attempt to classify B- and K-decays from the point
of view of theoretical cleanliness and their importance.

Whereas section 8 presents some future visions of the field of weak decays,
section 9 is a brief summary.
Finally section 10 contains some remarks on "Beauty 95".

\section{Theoretical Framework}
\subsection{Effective Field Theory Picture}
The basic framework for weak decays of hadrons containing u, d, s, c and
b quarks is the effective field theory relevant for scales $ \mu \ll
M_W, M_Z , m_t$. This framework
brings in local operators which govern ``effectively''
the transitions in question. From the point of view of the decaying
hadrons containing the lightest five quarks this is the only correct
picture we know and also the most efficient one for studying the
presence of QCD. Furthermore it represents the generalization of the Fermi
theory as formulated by Feynman and Gell-Mann \cite{GF}
almost forty years ago.

Indeed the simplest effective hamiltonian without QCD effects which
one would find from the first diagram
of fig. 1 is
\begin{equation}\label{Heff}
H^{o}_{eff} = \frac{G_F}{\sqrt 2} V_{cb} V_{cs}^*
(\bar c b)_{V-A}(\bar s c)_{V-A}
\end{equation}
where $G_F$ is the Fermi constant, $V_{ij}$ are the relevant CKM
factors and
\begin{equation}\label{OP}
(\bar c b )_{V-A}(\bar s c)_{V-A}\equiv
(\bar c \gamma_{\mu} (1-\gamma_5) b)
(\bar s \gamma_{\mu} (1-\gamma_5) c)=Q_2
\end{equation}
is a $(V-A)\cdot (V-A)$ current-current local operator denoted usually
by $Q_2$. The situation in the standard model is however more complicated
because of the presence of additional interactions which effectively
generate new operators. These are in particular the gluon, photon and
$Z^0$-boson exchanges and internal top contributions. Some of the elementary
interactions of this type are shown in fig. 1. Consequently the relevant
effective hamiltonian for B-meson decays involves generally several
operators $Q_i$ with various colour and Dirac structures which are different
from $Q_2$. Moreover each operator is multiplied by a calculable
coefficient $C_i(\mu)$:
\begin{equation}\label{HOPE}
H_{eff} = \frac{G_F}{\sqrt 2} V_{CKM} \sum_i
   C_i (\mu) Q_i
\end{equation}
 In this connection it should be mentioned that the usual Feynman
diagram drawings of the type shown in fig. 1 containing full  W-propagators,
$ Z^0-$propagators and top-quark propagators represent
really the happening at scales $ {\cal O}(M_W)$ whereas the true picture
of a decaying hadron is more correctly described by effective vertices
which are represented by the local operators in question.
Thus, whereas at scales $ {\cal O}(M_W) $ we deal with
the full six-quark theory
containing the photon, weak gauge bosons and gluons,
at scales ${\cal O}(1 GeV)$  the
relevant effective theory contains only three light quarks u, d and s,
gluons and the photon. At intermediate energy scales, $\mu={\cal O}(m_b)$
and $\mu={\cal O}(m_c)$,
relevant for beauty and charm decays effective five-quark
and effective four-quark theories have to be considered respectively.

The usual procedure then is to start at a high energy scale
${\cal O}(M_W) $
and consecutively integrate out the heavy degrees of freedom (heavy with
respect to the relevant scale $ \mu $) from explicitly appearing in the
theory. The word ``explicitly'' is very essential here. The heavy fields
did not disappear. Their effects are merely hidden in the effective
gauge coupling constants, running masses and most importantly in the
coefficients describing the ``effective'' strength of the operators at a
given scale $\mu $, the Wilson coefficient functions $C_i(\mu)$.

\subsection{OPE and Renormalization Group}
The Operator Product Expansion (OPE) combined with the renormalization group
approach can be regarded as a mathematical formulation of the picture
outlined above.
In this framework the amplitude for an {\it exclusive} decay $M\to F$
is written as
\begin{equation}\label{OPE}
 A(M \to F) = \frac{G_F}{\sqrt 2} V_{CKM} \sum_i
   C_i (\mu) \langle F \mid Q_i (\mu) \mid M \rangle
\end{equation}
where $M$ stands for the decaying meson, $F$ for a given final state and
$V_{CKM}$ denotes the relevant $CKM$ factor.
$ Q_i(\mu) $ denote
the local operators generated by QCD and electroweak interactions.
$ C_i(\mu) $ stand for the Wilson
coefficient functions (c-numbers).
The scale $ \mu $ separates the physics contributions in the ``short
distance'' contributions (corresponding to scales higher than $\mu $)
contained in $ C_i(\mu) $ and the ``long distance'' contributions
(scales lower than $ \mu $) contained in
$\langle F \mid Q_i (\mu) \mid M \rangle $.
 By evolving the scale from $ \mu ={\cal O}(M_W) $ down to
lower values of $ \mu $ one transforms the physics information
at scales higher
than $ \mu $ from the hadronic matrix elements into $ C_i(\mu) $. Since no
information is lost this way the full amplitude cannot depend on $ \mu $.
This is the essence of the renormalization group equations which govern the
evolution $ (\mu -dependence) $ of $ C_i(\mu) $. This $ \mu $-dependence
must be cancelled by the one present in $\langle  Q_i (\mu)\rangle $.
 It should be
stressed, however, that this cancellation generally involves many
operators due to the operator mixing under renormalization.

The general expression for $ C_i(\mu) $ is given by:
\begin{equation}\label{CV}
 \vec C(\mu) = \hat U(\mu,M_W) \vec C(M_W)
\end{equation}
where $ \vec C $ is a column vector built out of $ C_i $'s.
$\vec C(M_W)$ are the initial conditions which depend on the
short distance physics at high energy scales.
In particular they depend on $m_t$.
$ \hat U(\mu,M_W) $, the evolution matrix,
is given as follows:
\begin{equation}\label{UM}
 \hat U(\mu,M_W) = T_g \exp \left[
   \int_{g(M_W)}^{g(\mu)}{dg' \frac{\hat\gamma^T(g')}{\beta(g')}}\right]
\end{equation}
with $g$ denoting the QCD effective coupling constant. $ \beta(g) $
governs the evolution of $g$ and $ \hat\gamma $ is the anomalous dimension
matrix of the operators involved. The structure of this equation
makes it clear that the renormalization group approach goes
 beyond the usual perturbation theory.
Indeed $ \hat  U(\mu,M_W) $ sums automatically large logarithms
$ \log M_W/\mu $ which appear for $ \mu<<M_W $. In the so-called leading
logarithmic approximation (LO) terms $ (g^2\log M_W/\mu)^n $ are summed.
The next-to-leading logarithmic correction (NLO) to this result involves
summation of terms $ (g^2)^n (\log M_W/\mu)^{n-1} $ and so on.
This hierarchic structure gives the renormalization group improved
perturbation theory.

As an example let us consider only QCD effects and the case of a single
operator so that (\ref{CV}) reduces to

\begin{equation}\label{CC}
  C(\mu) =  U(\mu,M_W)  C(M_W)
\end{equation}
with $C(\mu)$ denoting the coefficient of the operator in question.
 Keeping the first two terms in the expansions of
 $\gamma(g)$ and $\beta(g)$ in powers of $g$:
\begin{equation}
\gamma (g) = \gamma^{(0)} \frac{\alpha_s}{4\pi} + \gamma^{(1)}
\frac{\alpha^2_s}{16\pi^2}
\quad , \quad
 \beta (g) = - \beta_0 \frac{g^3}{16\pi^2} - \beta_1
\frac{g^5}{(16\pi^2)^2}
\end{equation}
and inserting these expansions into (\ref{UM}) gives:
\begin{equation}\label{UMNLO}
 U (\mu, M_W) = \Biggl\lbrack 1 + {{\alpha_s (\mu)}\over{4\pi}} J
\Biggl\rbrack \Biggl\lbrack {{\alpha_s (M_W)}\over{\alpha_s (\mu)}}
\Biggl\rbrack^P \Biggl\lbrack 1 - {{\alpha_s (M_W)}\over{4\pi}} J
\Biggl\rbrack
\end{equation}
where
\begin{equation}
P = {{\gamma^{(0)}}\over{2\beta_0}},~~~~~~~~~ J = {{P}\over{\beta_0}}
\beta_1 - {{\gamma^{(1)}}\over{2\beta_0}}.
\end{equation}
General
formulae for $ \hat U (\mu, M_W) $ in the case of operator mixing and
valid also for electroweak effects can be found in \cite{BJLW}.
The leading
logarithmic approximation corresponds to setting $ J = 0 $ in (\ref{UMNLO}).

Clearly in order to calculate the full amplitude in (\ref{OPE}) also
the matrix elements $\langle F \mid Q_i (\mu) \mid M \rangle$ have to
be evaluated. Since they involve long distance contributions one is
forced in this case to use non-perturbative methods such as
lattice calculations, the
$1/N$ expansion, QCD sum rules or chiral perturbation theory. In the
case of semi-leptonic B meson decays also the Heavy Quark Effective Theory
(HQET) \cite{NEU} turns out to be a useful tool.
In HQET the matrix elements are
evaluated approximately in an expansion in $1/m_b$. Needless
to say all these non-perturbative methods have some limitations.
 Consequently the dominant theoretical
uncertainties in the decay amplitudes reside in
the matrix elements of $Q_i$.

\vspace{17.7cm}
\centerline{Fig. 1}

\subsection{Classification of Operators}
 Below we give six classes of operators which play the
dominant role in the phenomenology of weak decays. Typical diagrams in
the full theory from which these operators originate are indicated
 and shown in Fig. 1. The cross in Fig. 1d indicates
that magnetic penguins originate from the mass-term on the external
line in the usual QCD or QED penguin diagrams.
The six classes are given as follows ($\alpha$ and $\beta$ are colour
indices):

{\bf Current--Current (Fig. 1a):}
\begin{equation}\label{O1}
Q_1 = (\bar c_{\alpha} b_{\beta})_{V-A}\;(\bar s_{\beta} c_{\alpha})_{V-A}
{}~~~~~~Q_2 = (\bar c b)_{V-A}\;(\bar s c)_{V-A}
\end{equation}

{\bf QCD--Penguins (Fig. 1b):}
\begin{equation}\label{O2}
Q_3 = (\bar s b)_{V-A}\sum_{q=u,d,s,c}(\bar qq)_{V-A}~~~~~~
 Q_4 = (\bar s_{\alpha} b_{\beta})_{V-A}\sum_{q=u,d,s,c}(\bar q_{\beta}
       q_{\alpha})_{V-A}
\end{equation}
\begin{equation}\label{O3}
 Q_5 = (\bar s b)_{V-A} \sum_{q=u,d,s,c}(\bar qq)_{V+A}~~~~~
 Q_6 = (\bar s_{\alpha} b_{\beta})_{V-A}\sum_{q=u,d,s,c}
       (\bar q_{\beta} q_{\alpha})_{V+A}
\end{equation}

{\bf Electroweak--Penguins (Fig. 1c):}
\begin{equation}\label{O4}
Q_7 = {3\over 2}\;(\bar s b)_{V-A}\sum_{q=u,d,s,c}e_q\;(\bar qq)_{V+A}
{}~~~~~ Q_8 = {3\over2}\;(\bar s_{\alpha} b_{\beta})_{V-A}\sum_{q=u,d,s,c}e_q
        (\bar q_{\beta} q_{\alpha})_{V+A}
\end{equation}
\begin{equation}\label{O5}
 Q_9 = {3\over 2}\;(\bar s b)_{V-A}\sum_{q=u,d,s,c}e_q(\bar q q)_{V-A}
{}~~~~~Q_{10} ={3\over 2}\;
(\bar s_{\alpha} b_{\beta})_{V-A}\sum_{q=u,d,s,c}e_q\;
       (\bar q_{\beta}q_{\alpha})_{V-A}
\end{equation}

{\bf Magnetic--Penguins (Fig. 1d):}
\begin{equation}\label{O6}
Q_{7\gamma}  =  \frac{e}{8\pi^2} m_b \bar{s}_\alpha \sigma^{\mu\nu}
          (1+\gamma_5) b_\alpha F_{\mu\nu}\qquad
Q_{8G}     =  \frac{g}{8\pi^2} m_b \bar{s}_\alpha \sigma^{\mu\nu}
   (1+\gamma_5)T^a_{\alpha\beta} b_\beta G^a_{\mu\nu}
\end{equation}

{\bf $\Delta S = 2 $ and $ \Delta B = 2 $ Operators (Fig. 3e):}
\begin{equation}\label{O7}
Q(\Delta S = 2)  = (\bar s d)_{V-A} (\bar s d)_{V-A}~~~~~
 Q(\Delta B = 2)  = (\bar b d)_{V-A} (\bar b d)_{V-A}
\end{equation}

{\bf Semi--Leptonic Operators (Fig. 1f):}
\begin{equation}\label{9V}
Q_{9V}  = (\bar s b  )_{V-A} (\bar e e)_{V}~~~~~
Q_{10A}  = (\bar s b )_{V-A} (\bar e e)_{A}
\end{equation}

\subsection{Towards Phenomenology}
The rather formal expression for the decay amplitudes given in
(\ref{OPE}) can always be cast in the form \cite{PBE}:
\begin{equation}\label{PBEE}
A(M\to F)=\sum_i B_i V_{CKM}^{i} \eta^{i}_{QCD} F_i(m_t,m_c)
\end{equation}
which is more useful for phenomenology. In writing (\ref{PBEE})
we have generalized (\ref{OPE}) to include several CKM factors.
$F_i(m_t,m_c)$, the Inami-Lim functions,
 result from the evaluation of loop diagrams with
internal top and charm exchanges (see fig. 1) and may also depend
solely on $m_t$ or $m_c$. In the case of current-current operators
the $F_i$ are mass independent. The factors $\eta^{i}_{QCD}$ summarize
the QCD corrections which can be calculated by formal methods
discussed above. Finally $B_i$ stand for nonperturbative factors
related to the hadronic matrix elements of the contributing
operators: the main theoretical uncertainty in the whole enterprise.
In leptonic and semi-leptonic decays for which only the matrix elements
of weak currents are needed,
the non-perturbative $B$-factors can fortunately be determined from
leading tree level decays reducing
or removing the non-perurbative uncertainty. In non-leptonic
decays this is generally not possible and we have to rely on
existing non-perturbative methods. A well known example of a
$B_i$-factor is the renormalization group invariant parameter
$B_K$ \cite{BSS} defined by
\begin{equation}\label{bk}
B_K=B_K(\mu)\left[\alpha_s(\mu)\right]^{-2/9}
\qquad
\langle \bar K^{o}\mid Q(\Delta S=2)\mid K^{o}\rangle=
\frac{8}{3} B_K(\mu)F_K^2 m_K^2
\end{equation}
In order to simplify the presentation we have omitted the NLO
correction in $B_K$.
$B_K$ plays an important role in the phenomenology of CP violation
in $K \to \pi\pi$.

If one is interested in exhibiting the $m_t$ dependence, the expression
(\ref{PBEE}) can further be rewritten \cite{PBE}
as a linear combination of {\it seven} basic, universal (process
independent) $ m_t$--dependent functions $ F_r(x_t) $ with coefficients
$ P_r $ characteristic for the decay considered:
\begin{equation}\label{PE}
A(M \to F)=P_o+\sum_r P_r F_r(x_t)
\quad, \quad x_t=\frac{m_t^2}{M_W^2}
\end{equation}
where the sum is over all possible functions contributing to a given
amplitude.
The first term is related to
contributions involving other quarks, in particular the charm quark.
This is the  Penguin--Box Expansion (PBE) mentoned previously.

Equation (\ref{PE}) can be derived from the usual OPE by setting
$ \mu = M_W $
and decomposing properly the $ m_t $ dependence of different $ C_i(M_W) $
into the basic functions $ F_r(x_t) $ which result
from the diagrams of fig. 1.
The latter have been evaluated by Inami and Lim \cite{IL}. Since the
$ \gamma$--penguins, $ Z^0$--penguins and $ \Delta S = 1 $ box diagrams
are gauge dependent,
 the idea of \cite{PBE} was to consider a set of functions
$ F_r(x_t) $ corresponding to gauge independent combinations. The
approximate expressions for $ F_r(x_t) $
are given as follows:
\begin{equation}\label{PBE1}
 S(x_t)=0.784~x_t^{0.76},~~~~X(x_t)=0.660~x_t^{0.575},
\end{equation}
\begin{equation}\label{PBE2}
 Y(x_t)=0.315~x_t^{0.78},\quad Z(x_t)=0.175~x_t^{0.93},  \quad
   E(x_t)=0.564~x_t^{-0.51},
\end{equation}
\begin{equation}
 D'(x_t)=0.244~x_t^{0.30},~~~~~~~~~~E'(x_t)=0.145~x_t^{0.19}.
\end{equation}
In the range $150~GeV \le m_t\le 200~GeV$ these approximations reproduce the
exact expressions to an accuracy beter than 1\%.
$ S(x_t) $ results from the $ \Delta S = 2 $ or $ \Delta B = 2 $ box diagram,
$ E(x_t) $ from
the gluon penguin, $ D'(x_t) $ from the magnetic $\gamma$--penguin and
$ E'(x_t) $ from the magnetic gluon penguin. $ X(x_t) $ and $ Y (x_t) $ are
linear combinations of the V--A component of $ Z^0$--penguin and
box--diagrams with final quarks or leptons having  $ T_3 $ (weak isospin)
equal to 1/2 and -- 1/2 respectively. Finally $ Z(x_t) $ is a linear
combination of the vector component of the $ Z^0$--penguin and the
$\gamma$--penguin.

The coefficients $ P_r $ include the physics from scales
$ 0 \le \mu \le M_W $ and also the CKM factors which have been shown
explicitly in (\ref{PBEE}). Generally in a given decay
several of the coefficients vanish so that the corresponding amplitude
depends only on one or a few functions.

\begin{table}
\begin{center}
\begin{tabular}{|l|l|}
\hline
\bf \phantom{XXXXXX} Decay & \bf \phantom{XX} Reference~~~ \\
\hline
\hline
\multicolumn{2}{|c|}{$\Delta F=1$ Decays} \\
\hline
current-current operators     & \cite{ALTA,BW} \\
QCD penguin operators         & \cite{BJLW1,BJLW,ROMA1,ROMA2} \\
electroweak penguin operators & \cite{BJLW2,BJLW,ROMA1,ROMA2} \\
magnetic penguin operators    & \cite{MisMu:94}  \\
$Br(B)_{SL}$                  & \cite{ALTA,Buch:93,Bagan} \\
inclusive $\Delta S=1$ decays       & \cite{JP} \\
\hline
\multicolumn{2}{|c|}{Particle-Antiparticle Mixing} \\
\hline
$\eta_1$                   & \cite{HNa} \\
$\eta_2,~\eta_B$           & \cite{BJW} \\
$\eta_3$                   & \cite{HNb} \\
\hline
\multicolumn{2}{|c|}{Rare K- and B-Meson Decays} \\
\hline
$K^0_L \rightarrow \pi^0\nu\bar{\nu}$, $B \rightarrow l^+l^-$,
$B \rightarrow X_{\rm s}\nu\bar{\nu}$ & \cite{BB1,BB2} \\
$K^+   \rightarrow \pi^+\nu\bar{\nu}$, $K_L \rightarrow \mu^+\mu^-$
                                      & \cite{BB3} \\
$K^+\to\pi^+\mu\bar\mu$               & \cite{BB5} \\
$K_L \rightarrow \pi^0e^+e^-$         & \cite{BLMM} \\
$B\rightarrow X_s e^+e^-$           & \cite{Mis:94,BuMu:94} \\
\hline
\end{tabular}
\end{center}
\centerline{}
\caption{References to NLO Calculations}
\label{TAB1}
\end{table}

\subsection{Inclusive Decays}
So far we have discussed only {\it exclusive} decays. During the
recent years considerable progress has been made for inclusive
decays of heavy mesons. The starting point is again the effective
hamiltonian in (\ref{HOPE}) which includes the short distance QCD
effects in $C_i(\mu)$. The actual decay described by the operators
$Q_i$ is then calculated in the spectator model corrected for
additional virtual and real gluon corrections.
Support for this approximation
comes from heavy quark ( $1/m_b $) expansions (HQE).
Indeed the spectator
model has been shown to correspond to the leading order approximation
in the $1/m_b$ expansion.
The next corrections appear at the ${\cal O}(1/m_b^2)$
level. The latter terms have been studied by several authors
\cite{Chay,Bj,Bigi} with the result that they affect various
branching ratios by less than $10\%$ and often by only a few percent.
There is a vast literature on this subject and I can only refer here to
a few papers \cite{Bigi,Mannel} where further references can be found.
Of particular importance for this field was also the issue of the
renormalons which are nicely discussed in \cite{Beneke,Braun}.

\subsection{Weak Decays Beyond Leading Logarithms}
Until 1989 most of the calculations in the field of weak
decays were done in the leading logarithmic approximation.
An exception was the important work of Altarelli et al. \cite{ALTA}
who in 1981 calculated NLO QCD corrections to the Wilson
coefficients of the current-current operators.

Today the effective hamiltonians for weak decays are
available at the next-to-leading level for the most important
and interesting cases due to a series of publications devoted
to this enterprise beginning with the work of Peter Weisz and myself
in 1989 \cite{BW}. The list of the existing calculations is given in
table \ref{TAB1}.
We will discuss some of the entries in this list below.
A detailed review
of the existing NLO calculations will appear soon \cite{BBL}.

Let us recall why NLO calculations are important for the
phenomenology of weak decays:

\begin{itemize}
\item The NLO is first of all necessary to test the validity of
the renormalization group improved perturbation theory.
\item Without going to NLO the QCD scale $\Lambda_{\overline{MS}}$
extracted from various high energy processes cannot be used
meaningfully in weak decays.
\item Due to the renormalization group invariance the physical
amplitudes do not depend on the scales $\mu$ present in $\alpha_s$
or in the running quark masses, in particular $m_t(\mu)$,
$m_b(\mu)$ and $m_c(\mu)$. However
in perturbation theory this property is broken through the truncation
of the perturbative series. Consequently one finds sizable scale
ambiguities in the leading order, which can be reduced considerably
by going to NLO.
\item In several cases the central issue of the top quark mass dependence
is strictly a NLO effect.
\end{itemize}

\section{Standard Analysis}
\subsection{The CKM Matrix and the Unitarity Triangle}
An important target of particle physics is the determination
 of the unitary $3\times 3$ Cabibbo-Kobayashi-Maskawa
matrix \cite{CAB,KM} which parametrizes the charged current interactions of
 quarks:
\begin{equation}\label{1j}
J^{cc}_{\mu}=(\bar u,\bar c,\bar t)_L\gamma_{\mu}
\left(\begin{array}{ccc}
V_{ud}&V_{us}&V_{ub}\\
V_{cd}&V_{cs}&V_{cb}\\
V_{td}&V_{ts}&V_{tb}
\end{array}\right)
\left(\begin{array}{c}
d \\ s \\ b
\end{array}\right)_L
\end{equation}
The CP violation in the standard model is supposed to arise
from a single phase in this matrix.
It is customary these days to express the CKM-matrix in
terms of four Wolfenstein parameters
\cite{WO} $(\lambda,A,\varrho,\eta)$
with $\lambda=\mid V_{us}\mid=0.22 $ playing the role of an expansion
parameter and $\eta$
representing the CP violating phase:
\begin{equation}\label{2.75}
V_{CKM}=
\left(\begin{array}{ccc}
1-{\lambda^2\over 2}&\lambda&A\lambda^3(\varrho-i\eta)\\ -\lambda&
1-{\lambda^2\over 2}&A\lambda^2\\ A\lambda^3(1-\varrho-i\eta)&-A\lambda^2&
1\end{array}\right)
+O(\lambda^4)
\end{equation}
Because of the
smallness of $\lambda$ and the fact that for each element
the expansion parameter is actually
$\lambda^2$, it is sufficient to keep only the first few terms
in this expansion.

\vspace{7.2cm}
\centerline{Fig. 2}

Following \cite{BLO} one can define the parameters
$(\lambda, A, \varrho, \eta)$ through
\begin{equation}\label{wop}
s_{12}\equiv\lambda \qquad s_{23}\equiv A \lambda^2 \qquad
s_{13} e^{-i\delta}\equiv A \lambda^3 (\varrho-i \eta)
\end{equation}
where $s_{ij}$ and $\delta$ enter the standard exact
parametrization \cite{PDG}  of the CKM
matrix. This specifies the higher orders terms in (\ref{2.75}).

The definition of $(\lambda,A,\varrho,\eta)$ given in (\ref{wop})
is useful because it allows to improve the accuracy of the
original Wolfenstein parametrization in an elegant manner. In
particular
\begin{equation}\label{CKM1}
V_{us}=\lambda \qquad V_{cb}=A\lambda^2
\end{equation}
\begin{equation}\label{CKM2}
V_{ub}=A\lambda^3(\varrho-i\eta)
\qquad
V_{td}=A\lambda^3(1-\bar\varrho-i\bar\eta)
\end{equation}
where
\begin{equation}\label{3}
\bar\varrho=\varrho (1-\frac{\lambda^2}{2})
\qquad
\bar\eta=\eta (1-\frac{\lambda^2}{2})
\end{equation}
turn out \cite{BLO} to be excellent approximations to the
exact expressions.

A useful geometrical representation of the CKM matrix is the unitarity
triangle obtained by using the unitarity relation
\begin{equation}\label{2.87h}
V_{ud}V_{ub}^* + V_{cd}V_{cb}^* + V_{td}V_{tb}^* =0,
\end{equation}
rescaling it by $\mid V_{cd}V_{cb}^\ast\mid=A \lambda^3$ and depicting
the result in the complex $(\bar\rho,\bar\eta)$ plane as shown
in fig. 2. The lenghts CB, CA and BA are equal respectively to 1,
\begin{equation}\label{2.94a}
R_b \equiv  \sqrt{\bar\varrho^2 +\bar\eta^2}
= (1-\frac{\lambda^2}{2})\frac{1}{\lambda}
\left| \frac{V_{ub}}{V_{cb}} \right|
\qquad
{\rm and}
\qquad
R_t \equiv \sqrt{(1-\bar\varrho)^2 +\bar\eta^2}
=\frac{1}{\lambda} \left| \frac{V_{td}}{V_{cb}} \right|.
\end{equation}

The triangle in fig. 2, $\mid V_{us}\mid$ and $\mid V_{cb}\mid$
give the full description of the CKM matrix.
Looking at the expressions for $R_b$ and $R_t$ we observe that within
the standard model the measurements of four CP
{\it conserving } decays sensitive to $\mid V_{us}\mid$, $\mid V_{ub}\mid$,
$\mid V_{cb}\mid $ and $\mid V_{td}\mid$ can tell us whether CP violation
($\eta \not= 0$) is predicted in the standard model.
This is a very remarkable property of
the Kobayashi-Maskawa picture of CP violation: quark mixing and CP violation
are closely related to each other.

There is of course the very important question whether the KM picture
of CP violation is correct and more generally whether the standard
model offers a correct description of weak decays of hadrons. In order
to answer these important questions it is essential to calculate as
many branching ratios as possible, measure them experimentally and
check if they all can be described by the same set of the parameters
$(\lambda,A,\varrho,\eta)$. In the language of the unitarity triangle
this means that the various curves in the $(\bar\varrho,\bar\eta)$ plane
extracted from different decays should cross each other at a single point.
Moreover the angles $(\alpha,\beta,\gamma)$ in the
resulting triangle should agree with those extracted one day from
CP-asymmetries in B-decays. More about this below.

\subsection{Basic Formulae}

At present there is still a rather limited knowledge of the shape of
the unitarity triangle. The standard analysis using the available
experimental and theoretical information proceeds essentially in four
steps:

{\bf Step 1:}

{}From  $b\to c$ transition in inclusive and exclusive B meson decays
one finds $\vcb$ and consequently the scale of UT:
\begin{equation}
\vcb\quad =>\quad\lambda \vcb= \lambda^3 A
\end{equation}

{\bf Step 2:}

{}From  $b\to u$ transition in inclusive B meson decays
one finds $\vub$ and consequently the side $CA=R_b$ of UT:
\begin{equation}
\vub \quad=> \quad R_b
\end{equation}

{\bf Step 3:}

{}From the observed indirect CP violation in $K \to \pi\pi$ described
experimentally by the parameter $\varepsilon_K$ and theoretically
by the imaginary part of the relevant box diagram in fig. 1e one
derives the constraint:
\begin{equation}\label{100}
\bar\eta \left[(1-\bar\varrho) A^2 \eta_2 S(x_t)
+ P_0(\varepsilon) \right] A^2 B_K = 0.226
\end{equation}
where
\begin{equation}\label{102}
P_0(\varepsilon) =
\left[ \eta_3 S(x_c,x_t) - \eta_1 x_c \right] \frac{1}{\lambda^4}
\end{equation}
Equation (\ref{100}) specifies
a hyperbola in the $(\bar \varrho, \bar\eta)$
plane. Here $S(x_t)$ is the function given in (\ref{PBE1}), $B_K$
is the non-perturbative parameter defined in (\ref{bk}) and $\eta_2$
is the QCD factor in the box diagrams with two top quark exchanges.
Finally $P_0(\varepsilon)=0.31\pm0.02$ summarizes the contributions
of box diagrams with two charm quark exchanges and the mixed
charm-top exchanges. $P_0(\varepsilon)$ depends very weakly on $m_t$ and
its range given above corresponds to $155~GeV \le m_t \le 185~GeV$.
The QCD factors $\eta_1$ for the (cc) contribution and $\eta_3$ for
the (ct) contribution are included in this calculation as seen
in (\ref{102}).

{\bf Step 4:}

{}From the observed $B^0_d-\bar B^0_d$ mixing described experimentally
by the mixing parameter $x_d=\Delta M/\Gamma_B$
and theoretically by the relevant box diagram of Fig. 1e
the side $BA=R_t$ of the UT can be determined:
\begin{equation}\label{106}
 R_t = 1.52 \cdot \frac{R_0}{\sqrt{ S(x_t)}}=0.97 R_0.
\left [\frac{170~GeV}{\bar m_t(m_t)} \right ]^{0.76}
\end{equation}
where
\begin{equation}\label{107}
R_0 \equiv \sqrt{ \frac{x_d}{0.75}} \left[ \frac{200 MeV}{F_{B_d}
 \sqrt{B_{B_d}}}
\right] \left[ \frac{0.040}{\vcb} \right]
\left[\frac{1.6\,ps}{\tau_B}\right]^{0.5}
 \sqrt{ \frac{0.55}{\eta_B}}
\end{equation}
This allows to determine $\vtd$:
\begin{equation}\label{VT}
\mid V_{td} \mid=
8.6\cdot 10^{-3}\left [
\frac{200~MeV}{\sqrt{B_B}F_B}\right ]
\left [\frac{170~GeV}{\bar m_t(m_t)} \right ]^{0.76}
\left [\frac{x_d}{0.75} \right ]^{0.5}
\left [\frac{1.6~ps}{\tau_B} \right ]^{0.5}
\sqrt{ \frac{0.55}{\eta_B}}
\end{equation}

Here $\tau_B$ is the B-meson life-time,
$\eta_B$ is the QCD factor analogous to $\eta_2$,
$F_{B_d}$ is the B-meson decay constant and $B_{B_d}$
denotes a non-perturbative
parameter analogous to $B_K$.
Note that whereas $R_t$ depends on $\vcb$, the determination of
$\vtd$ by means of $x_d$ is free from this dependence.

Before discussing the values of the input parameters in the basic
formulae given above we would like to make a few important messages
for  UT-practitioners.

\subsection{Messages for  UT Practitioners}

\subsubsection{\bf Message 1}

The parameter $m_t$, the top quark mass, used in weak decays is not
equal to the one used in the electroweak precision studies at LEP or
SLD. In the latter investigations the so-called pole mass is used,
whereas in all the NLO calculations listed in table 1 $m_t$ refers
to the running current top quark mass normalized at $\mu=m_t$:
$\bar m_t(m_t)$. One has
\begin{equation}
m_t^{Pole}=\bar m_t(m_t)
\left[ 1+\frac{4}{3}\frac{\alpha_s(m_t)}{\pi}\right]
\end{equation}
so that for $m_t={\cal O}(170~GeV)$, $\bar m_t(m_t)$ is typically
by $8~GeV$ smaller than $m_t^{Pole}$. This difference will matter in
a few years. There is also an interesting question: What is the meaning
of $m_t$ quoted by CDF \cite{CDF} and D0 \cite{D0}?
My feeling is that this is unclear at
present. I will assume, as most people do, that this is the pole mass
and consequently use in weak decays
\begin{equation}
m_t\equiv\bar m_t(m_t)=(170\pm15)~GeV
\end{equation}
In principle an error of $11~GeV$ could be used. In view of the
uncertainty in the definition of $m_t$ I prefer to be conservative.
Let us hope that the issue of the definition of $m_t$ at Fermilab
will be clarified soon.

\subsubsection{\bf Message 2}

When using numerical values for $m_t$, $B_K$, $B_B$ and the QCD factors
$\eta_i$, care must be taken that they are used consistently. This
unfortunately is not always the case. As an example let us consider
the theoretical expression for $x_d$ which reads
\begin{equation}\label{x_d}
x_d= C_B F_B^2 B_B(\mu_B) \left[\alpha_s(\mu_B)\right]^{-6/25}
     \eta_B(\mu_t,\bar m_t(\mu_t)) S(\bar m_t(\mu_t))
\end{equation}
with $C_B$ being a numerical constant. Here $S$ stands for the box
diagram function of (\ref{PBE1}) and $\eta_B$ represents the QCD
corrections to the relevant box diagrams. $B_B(\mu_B)$ is defined in
analogy to $B_K(\mu)$ in (\ref{bk}). Two relevant scales are
$\mu_B={\cal O}(m_b)$ and $\mu_t={\cal O}(m_t)$ which according to
the rules of the renormalization group game can be chosen for instance
in the ranges $2.5~GeV\le \mu_B \le 10~GeV$ and
$100~GeV \le \mu_t \le 300~GeV$ respectively. $\mu_B$ is the scale at
which the relevant $\Delta B=2$ operator is normalized, $\mu_t$ is
the scale at which $m_t$ is defined. Clearly $x_d$ cannot depend on
$\mu_B$ and $\mu_t$. Combining the explicit $\alpha_s$ factor in (\ref{x_d})
with $B_B(\mu_B)$ as in (\ref{bk}) and introducing the renormalization group
invariant $B_B$ removes $\mu_B$ from phenomenological expressions like
(\ref{VT}). On the other hand the $\mu_t$ dependence cancells between the last
two terms as demonstrated explicitly in \cite{BJW}.
To this end the NLO calculation
for $\eta_B$ is essential. Otherwise $x_d$ shows a sizable $\mu_t$
dependence. It turns out that for a choice $\mu_t=m_t$, $\eta_B$ and
similarly $\eta_2$ in (\ref{100}) are practically independent of $m_t$. This
is convenient and has been adopted in \cite{BJW} and in subsequent
NLO calculations.
Then $\eta_B=0.55$ and $\eta_2=0.57$ independent of $m_t$.

In "old days" the explicit $\alpha_s$ factor in (\ref{x_d}) has been
combined with $\eta_B$ to give the corresponding $\mu_B$ dependent
QCD factor as high as 0.85. This change is compensated by
$B_B(\mu_B)<B_B$. In view of the fact that all present "non-perturbative
researchers" quote $B_B$ and $B_K$, it is important that this older
definition is abandoned.

Similar messages apply to $\eta_i$ in the case of $\varepsilon_K$.

\subsubsection{\bf Message 3}

It is sometimes stated in the literature that the QCD factors $\eta_B$
for $B^0_d-\bar B^0_d$  and $B^0_s-\bar B^0_s$ mixings are "expected"
to be equal to each other. There is nothing to expect here. They are
equal. Indeed, $\eta_B$ resulting from short distance QCD calculations is
independent of whether $B^0_d-\bar B^0_d$  or $B^0_s-\bar B^0_s$ is considered.
Consequently the ratio $x_d/x_s$ is independent of
$m_t$ and short distance QCD corrections.
The only difference in these two mixings arises through different CKM
factors, life-times
and through different hadronic matrix elements of the relevant $\Delta B=2$
operators which corresponds to $F_{B_s}\not=F_{B_d}$ and
$B_{B_s}\not=B_{B_d}$. This is explicitly summarized in (\ref{107b}).

\subsection{Input Parameters}
Let us next discuss the input parameters needed to perform the
standard analysis.

\subsubsection{\bf $\mid V_{cb} \mid$}

During the last two years there has been a considerable progress
done by experimentalists \cite{PAT} and theorists in the extraction of
$\mid V_{cb} \mid$ from exclusive and inclusive decays. In
particular I would like to mention important papers by
Neubert \cite{Neubert}, Shifman, Uraltsev and Vainshtein \cite{SUV}, and
Ball, Benecke and Braun \cite{Braun}
on the basis of which one is entitled to use ( for $\tau_B=1.6~ps$):
\begin{equation}\label{2}
\mid V_{cb} \mid=0.040\pm0.003\quad =>\quad A=0.82\pm 0.06
\end{equation}

\subsubsection{\bf $\left| V_{ub}/V_{cb} \right|$}

Here the situation is much worse and the value
\begin{equation}\label{2.94}
\left| \frac{V_{ub}}{V_{cb}} \right|=0.08\pm0.02
\end{equation}
quoted by Particle Data \cite{PDG} appears to be still valid.
There is a hope that the error could be reduced by a factor of 2 to 4
in the coming years both due to the theory \cite{Braun,URAL} and
the recent CLEO measurements of the exclusive semileptonic decays
$B \to (\pi,\varrho)l\nu_l$ \cite{CLEO3}.

\subsubsection{\bf $\left| V_{ub}/V_{cb} \right|$
and $\left| V_{cb} \right|$}

The values in (\ref{2}) and (\ref{2.94}) are not correlated with
each other. On the other hand such a correlation is present in
the analysis of the CP violating parameter $\varepsilon_K$ which
is roughly proportional to the fourth power of $\left| V_{cb}\right|$
and linear in $\left|V_{ub}/V_{cb} \right|$. It follows
that not all values in (\ref{2}) and (\ref{2.94}) are simultaneously
consistent with the observed value of $\varepsilon_K$. This is indirectly
seen in \cite{AB} and has been more explicitly emphasized this year by
Herrlich and Nierste \cite{HNb} and in \cite{BBL}.
Updating and rewriting the analytic
lower bound on $m_t$ from $\varepsilon_K$ \cite{AB} one finds \cite{BBL}

\begin{equation}
\left| \frac{V_{ub}}{V_{cb}} \right|_{min}=
\frac{0.225}{B_K A^2(2 x_t^{0.76}A^2+1.4)}
\end{equation}

This bound is shown as a function of $\mid V_{cb} \mid$ for different
values of $B_K$ and $m_t=185~GeV$ in fig.3. We observe that simultaneously
small values of $\left| V_{ub}/V_{cb} \right|$ and $\left| V_{cb} \right|$
although still consistent with (\ref{2}) and (\ref{2.94}), are not allowed
by the size of the indirect CP violation observed in $K \to \pi\pi$.

\vspace{10.35cm}
\centerline{Fig. 3}

\subsubsection{\bf $\eta_i$ and $\eta_B$}
The NLO values for $\eta_2$ and $\eta_B$ have been known already
for some time \cite{BJW}:
\begin{equation}\label{KNLO1}
 \eta_2=0.57\pm0.01
\qquad
\eta_B=0.55\pm0.01
\end{equation}
The NLO values for $\eta_1$ and $\eta_3$ are relatively new with
the calculation of $\eta_3$ completed by Herrlich and Nierste
this summer \cite{HNa,HNb}:
\begin{equation}\label{KNLO2}
\eta_1=1.38\pm 0.20\qquad  \eta_3=0.47\pm0.04
\end{equation}

The quoted errors reflect the remaining theoretical uncertainties due to
$\Lambda_{\overline{MS}}$ and the quark
masses. The references to the leading order calculations can be found in
\cite{BBL}.

\subsubsection{\bf $B_K$}

Concerning the parameter $B_K$, the most recent analyses
using the lattice methods
\cite{SH0,Ishizuka} ($B_K=0.83\pm 0.03$) and the $1/N$ approach
 of \cite{BBG0}
modified somewhat in \cite{Bijnens} give results in the ball park
of the $1/N$ result $B_K=0.70\pm 0.10$ obtained a long
time ago \cite{BBG0}. In particular the analysis of Bijnens and Prades
\cite{Bijnens} seems to have explained the difference between these values
for $B_K$ and the lower values obtained by using the QCD Hadronic Duality
approach \cite{Prades} ($B_K=0.39\pm 0.10$) or using the $SU(3)$ symmetry and
 PCAC
($B_K=1/3$) \cite{Donoghue}. This is gratifying because such low values for
$B_K$ would require $m_t>250~GeV$ in order to explain the experimental
value of $\varepsilon_K$ \cite{AB,BLO,HNb}. In our numerical analysis we
will use
\begin{equation}
B_K=0.75\pm 0.15
\end{equation}

\subsubsection{\bf $F_B$}

There is a vast literature on the lattice calculations of $F_B$. The
most recent results are somewhat lower than quoted a few years ago.
Based on a review by Chris Sachrajda \cite{Chris}, the recent extensive
study by Duncan et al. \cite{Duncan} and the analyses in \cite{Latt}
we conclude:
$F_{B_d}=(180\pm40)~MeV$. This together with the earlier result of
the European Collaboration for $B_{B_d}$, gives
$F_{B_d}\sqrt{B_{B_d}}=195\pm 45~MeV$.
The reduction of the error in this important quantity is desirable.
These results for $F_B$ are compatible with the results obtained using
QCD sum rules (e.g.\cite{QCDS}). An interesting upper bound
$F_{B_d}<195~MeV$ using QCD dispersion relations has also recently
been obtained \cite{BGL}.
In our numerical analysis we will use:
\begin{equation}
F_{B_d}\sqrt{B_{B_d}}=200\pm 40~MeV.
\end{equation}
At this point let us recall that the measurement of
$B^o_d-\bar B^o_d$ mixing parametrized by $x_d$ together
with  $B^o_s-\bar B^o_s$ mixing parametrized by $x_s$ allows to
determine $R_t$:
\begin{equation}\label{107b}
R_t = \frac{1}{\sqrt{R_{ds}}} \sqrt{\frac{x_d}{x_s}} \frac{1}{\lambda}
\qquad
R_{ds} = \frac{\tau_{B_d}}{\tau_{B_s}} \cdot \frac{m_{B_d}}{m_{B_s}}
\left[ \frac{F_{B_d} \sqrt{B_{B_d}}}{F_{B_s} \sqrt{B_{B_s}}} \right]^2
\end{equation}
where $R_{ds}$ summarizes SU(3)--flavour breaking effects.
Note that $m_t$ and $V_{cb}$ dependences have been eliminated this way
 and that $R_{ds}$
contains much smaller theoretical
uncertainties than the hadronic matrix elements in $x_d$ and $x_s$
separately.
Provided $x_d/x_s$ has been accurately measured a determination
of $R_t$ within $\pm 10\%$ should be possible. Indeed the most
recent lattice result \cite{Duncan} gives $F_{B_s}/F_{B_d}=1.22\pm0.04$.
A similar result has been obtained using QCD sum rules \cite{NAR}.
It would be useful to know $B_{B_s}/B_{B_d}$ with similar precision.
For $B_{B_s}=B_{B_d}$ I find $R_{ds}=0.66\pm 0.07$.

\subsubsection{\bf $\tau_B$ and $x_d$}

At this meeting the world averages for B-life-times have been
presented by Giuliana Rizzo \cite{Rizzo}:
\begin{equation}
\tau(B_d)=(1.57\pm0.05)~ps
\qquad
\tau(B^+)=(1.63\pm0.05)~ps
\qquad
\tau(B_s)=(1.58\pm0.10)~ps
\end{equation}
On the other hand the values for $x_d$ and $x_s$ have been summarized
by Achille Stocchi \cite{Stocchi}:
\begin{equation}
x_d=0.714\pm 0.043
\qquad
x_s>8.3~ (95\%~ C.L.)
\end{equation}
This is compatible with $x_d=0.73\pm 0.04$ of Aleksan \cite{Aleksan}.
In what follows I will use
\begin{equation}
\tau(B_d)=\tau(B_s)=1.6~ps
\qquad
x_d=0.75\pm0.06
\end{equation}
which is compatible with the above values. I prefer to use $x_d$
instead of $\Delta M$ because $x_d$ is dimensionless.

\subsubsection{\bf $m_t$}

Finally it is important to stress that the discovery of the top quark
\cite{CDF,D0} and its mass measurement had an important impact on
the field of rare decays and CP violation reducing considerably one
potential uncertainty. As we stated above, in loop induced  B
decays the relevant mass parameter is the running current quark mass.
 In this review we will
simply denote this mass by $m_t$ and use:
\begin{equation}
m_t\equiv\bar m_t(m_t)=(170\pm15)~GeV
\end{equation}

\vspace{11.2cm}
\centerline{Fig. 4}

\subsection{"Future" Input Parameters}
The "present" input parameters for the standard analysis are summarized
above. It is expected that the future will bring the reduction of
errors in the input parameters. For this reason I will also present
the results using:

\begin{equation}\label{200}
\begin{array}{rclrcl}
\left| V_{cb} \right| & = &  0.040 \pm 0.001 &
\mid V_{ub}/V_{cb} \mid & = & 0.08 \pm 0.01 \\
B_K & = & 0.75 \pm 0.05 & \sqrt{B_{B_d}} F_{B_d} & = & (200 \pm 10)~MeV \\
x_d & = & 0.75 \pm 0.03 & m_t & = & (170 \pm 5)~GeV \\
\end{array}
\end{equation}
with all other input parameters unchanged. It is plausible that these
reduced errors will be achieved at the end of this decade, although
the central values are not guaranteed.

\subsection{Output of a Standard Analysis}
In table 2 I show the ranges for various quantities of interest which
have been found using the sets of input parameters just discussed.
This analysis has been done in collaboration with Gerhard Buchalla and
Markus Lautenbacher. More details and more results can be found in
\cite{BBL}.
In fig. 4  I show the present (a) and the future (b) ranges for the upper
corner A of the UT. The solid lines correspond to $R_t$ from (\ref{107b})
using $R_{ds}=0.66$ and $x_s=10,15,25$ and $40$, respectively.
The allowed region has a typical "banana" shape which can be found
in many other analyses \cite{BLO,RUT,HNb,ALUT,PW}. The size of
the banana and its position depends on the assumed input
parameters and on the error analysis which varies from paper
to paper. The results in fig. 4 correspond to a simple independent
scanning of all parameters within one standard deviation.
Effectively such an approach is more conservative than using
Gaussian distributions as done in some papers quoted above.
Since all these results are selfexplanatory, let me move to other
topics.

\begin{table}
\begin{center}
\begin{tabular}{|c||c||c|}\hline
& Present & Future \\ \hline
$\mid V_{td}\mid/10^{-3}$ &$6.7 - 11.9$ &$ 7.8 -9.4$ \\ \hline
$\mid V_{ts}/V_{cb}\mid$ &$0.959 - 0.993$ &$0.975 - 0.986$  \\ \hline
$\sin(2\beta)$ &$0.33 - 0.80$ &$ 0.57 - 0.73 $ \\ \hline
$\sin(2\alpha)$ &$-0.86 - 1.0$ &$ -0.30 - 0.73 $ \\ \hline
$\sin(\gamma)$ &$0.61 - 1.0 $ &$ 0.85 - 1.0 $ \\ \hline
$x_s$ &$ 11.3 - 46.7$ &$ 19.5 - 29.4 $ \\ \hline
\end{tabular}
\end{center}
\centerline{}
\caption{ Output for "present" and "future" inputs.}
\label{TAB2}
\end{table}

\section{Three Topics}
\subsection{NLO Corrections to $B_{SL}$}
An important issue are the NLO corrections
to the non-leptonic width
for B-Mesons which is relevant for the theoretical prediction of
the inclusive semileptonic branching ratio:
\begin{equation}
B_{SL}=\frac{\Gamma(B\to Xe\nu)}{\Gamma_{SL}(B)+\Gamma_{NL}(B)}
\end{equation}
This calculation can be done within the spectator model corrected
for small non-perturbative corrections \cite{Bigi}
 and more important gluon
bremsstrahlung and virtual gluon corrections.
The calculation of NLO QCD corrections to $\Gamma_{NL}(B)$
has been done by Altarelli et al.~\cite{ALTA}
 in the
DRED renormalization scheme and by Buchalla \cite{Buch:93} in the HV scheme
neglecting the masses of the final quarks.
 The results of these papers agree with each other.

It is well known that the inclusion of QCD corrections in the
spectator model lowers $B_{SL}$ which otherwise
would be roughly $16\%$. Unfortunately the theoretical branching ratio
based on the QCD calculation of refs. \cite{ALTA,Buch:93}  turns out to be
 typically $B_{SL}=(12.5-13.5)\%$ \cite{AP:92}
whereas the experimental world average \cite{Browder} is
\begin{equation}\label{BEXP}
B^{exp}_{SL}=(10.4\pm 0.4)\%
\end{equation}
The inclusion of the leading non-perturbative correction
${\cal O}(1/m_b^2)$ lowers slightly the theoretical
prediction but gives only $\Delta_{NP} B_{SL}=-0.2\%$ \cite{Bigi}.
On the other hand mass effects in the QCD corrections to $B_{SL}$
seem to play an important role.
Bagan et al. \cite{Bagan} using partially the results of Hokim and Pham
\cite{Pham} have demonstrated that the inclusion
of mass effects in
the QCD calculations of refs.\cite{ALTA,Buch:93} (in particular in the
decay $b \to c\bar c s$ (see also \cite{Voloshin}) and taking into account
various renormalization
scale uncertainties improves the situation considerably.
Bagan et al. find \cite{Bagan}:
\begin{equation}
B_{SL}=(12.0\pm 1.4)\% \quad {\rm and} \quad \bar B_{SL}=(11.2\pm1.7)\%
\end{equation}
for the pole quark masses and $\overline{MS}$ masses respectively.
Within existing uncertainties, this result does not disagree
significantly with the experimental value, although it is still
somewhat on the high side.
\subsection{The $n_c$ Problem}
The number of charmed hadrons per B decay, $n_c$, is measured as
\cite{Browder}:
\begin{equation}\label{NEXP}
n_c=1.10\pm 0.06
\end{equation}
There appears to be a conflict between (\ref{BEXP}) and (\ref{NEXP})
as stressed by many authors \cite{nc,Bagan,Voloshin}. Indeed
the explanation of (\ref{BEXP}) requires substantial rate for
$b \to c\bar c s$ which in turn implies $n_c\approx 1.3$. Until
recently the error on the predicted $n_c$ was too large to draw
any definite conclusions. However recently, in an interesting analysis,
Buchalla, Dunietz and Yamamoto \cite{BDY} have considerably improved
this estimate in a hybrid approach which combines reliable
theoretical calculations with well measured quantities from
experiment. They find
\begin{equation}\label{NTH}
n_c=1.30\pm 0.05
\end{equation}
which is significantly larger than the experimental
value in (\ref{NEXP}). They give arguments why the current
experimental value of $n_c$ is underestimated and expect an increase
in the measured $n_c$ in the future. It will be interesting to
see whether this will turn out to be the case.
\subsection{Factorization}
An important issue is the question of factorization in non-leptonic
B-decays. This topic has been extensively discussed by Reinhold
R\"uckl \cite{RR}
at this workshop and I will only make a few remarks here.

In the strict factorization approach the matrix elements of the operators
$Q_i$ in (\ref{OPE}) are factored into the product of matrix elements
of quark currents. Since the resulting matrix elements of $Q_i$ are then
$\mu$ independent they cannot cancel the $\mu$ dependence
of the Wilson coefficients which in the factorization approch
are given by effective coefficients $a_i(\mu)$ \cite{STECH}.
The full amplitude
depends then on $\mu$ which is unphysical.
This $\mu$ dependence can only be cancelled
by non-factorizable contributions in the matrix elements. One
can try to argue that there is some "physical" value of $\mu$
at which factorization takes place. The problem is that the $a_i(\mu)$
depend generally also on the renormalization scheme, in particular on
the treatment of $\gamma_5$ in $D\not=4$ dimensions.
This dependence can again be cancelled only by non-factorizable
contributions.
A numerical analysis of the $\mu$ and scheme dependence of
$a_i(\mu)$  can be found in \cite{AB:95c}. It has been found
that $a_1(\mu)$ is almost independent both of $\mu$ and the
renormalization
scheme indicating that the factorization in so-called class 1
decays ("external W-exchange") could afterall be a good approximation.
This is in fact the class for which some plausible arguments
for factorization can be given \cite{Bjj}. In the case of $a_2(\mu)$ the
$\mu$ and scheme dependences are found to be very large implying
the importance of non-factorizable contributions in class 2
decays to which $B_d \to \psi K_S$ belongs. This can be demonstrated
using QCD sum rules as discussed by R\"uckl in these proceedings.

\begin{table}
\begin{center}
\begin{tabular}{|c||c|}\hline
 Deep Inelastic Scattering &  Non-Leptonic Decays \\ \hline \hline
Bjorken Scaling &  Factorization  \\ \hline
Scaling Violations (QCD Logs) &  Factorization Breakdown through $\mu$
 \\ \hline
 Scaling Violations (Higher Twists) & Final State Interactions
 \\ \hline
\end{tabular}
\end{center}
\centerline{}
\caption{ Analogy between DIS and Non-Leptonic Decays.}
\label{TAB3}
\end{table}

One should stress that there are still other non-factorizable
contributions like final state interactions. In a sense there is
here some analogy to deep-inelastic scattering. I show this in table 3.
Related issues are discussed by R\"uckl. In any case studies of
factorization and in particular the searches for patterns of factorization
breakdown in non-leptonic decays are very important and should
be pursued.

\section{Rare B Decays}
\subsection {$B\to X_s\gamma$}
The rare decay $B\to X_s\gamma$ plays an important role in the
present day phenomenology. It originates from the magnetic
$\gamma$-penguin of fig. 1d.
The effective hamiltonian for $B\to X_s\gamma$ at scales $\mu=O(m_b)$
is given by
\begin{equation} \label{Heff_at_mu}
{\cal H}_{eff}(b\to s\gamma) = - \frac{G_F}{\sqrt{2}} V_{ts}^* V_{tb}
\left[ \sum_{i=1}^6 C_i(\mu) Q_i + C_{7\gamma}(\mu) Q_{7\gamma}
+C_{8G}(\mu) Q_{8G} \right]
\end{equation}
where in view of $\mid V_{us}^*V_{ub} / V_{ts}^* V_{tb}\mid < 0.02$
 we have neglected the term proportional to $V_{us}^* V_{ub}$.
The magnetic $\gamma$-penguin is represented here by the operator
$Q_{7\gamma}$.

 The complicated structure of (\ref{Heff_at_mu}) makes
it clear that representing this decay simply by the diagram in fig.1d
misrepresents the true situation.
 Indeed the perturbative QCD effects involving in particular the
operator $Q_2$ are very important in this decay.
They are known
\cite{Bert,Desh} to enhance $B\to X_s\gamma$ in
the SM by 2--3
times, depending on the top quark mass. Since the first analyses
in \cite{Bert,Desh} a lot of progress has been made in calculating
the QCD effects beginning with the work in \cite{Grin,Odon}. We will
briefly summarize this progress.

A peculiar feature of the renormalization group analysis
in $B\to X_s\gamma$ is that the mixing under infinite renormalization
between
the set $(Q_1...Q_6)$ and the operators $(Q_{7\gamma},Q_{8G})$ vanishes
at the one-loop level. Consequently in order to calculate
the coefficients
$C_{7\gamma}(\mu)$ and $C_{8G}(\mu)$ in the leading logarithmic
approximation, two-loop calculations of ${\cal{O}}(e g^2_s)$
and ${\cal{O}}(g^3_s)$
are necessary. The corresponding NLO analysis requires the evaluation
of the mixing in question at the three-loop level.

At present, the coefficients $C_{7\gamma}$ and $C_{8G}$ are only known
in the leading logarithmic approximation.
However the peculiar feature of this decay mentioned above caused
that the first fully correct calculation of the leading  anomalous
dimension matrix has been obtained only in 1993 \cite{CFMRS:93,CFRS:94}.
It has been
confirmed subsequently in \cite{CCRV:94a,CCRV:94b,Mis:94}.
The NLO corrections are only partly known. The two-loop
mixing involving the operators
$Q_1.....Q_6$ and the two-loop mixing
in the sector $(Q_{7\gamma},Q_{8G})$ has been calculated in
\cite{ALTA,BW,BJLW1,BJLW,ROMA1,ROMA2}
and \cite{MisMu:94} respectively.
The calculation of the three loop mixing between
the set $(Q_1...Q_6)$ and the operators $(Q_{7\gamma},Q_{8G})$
 has not be done. The $O(\alpha_s)$
corrections to $C_{7\gamma}(M_W)$ and $C_{8G}(M_W)$ have been considered
in \cite{Yao1}. Gluon corrections to the matrix elements of magnetic
penguin operators have been calculated in \cite{AG1,AG2}.

The leading
logarithmic calculations
\cite{Grin,CFRS:94,CCRV:94a,Mis:94,AG1,BMMP:94}
   can be summarized in a compact form
as follows:
\begin{equation}\label{main}
\frac{Br(B \to X_s \gamma)}{Br(B \to X_c e \bar{\nu}_e)}
 =  \frac{|V_{ts}^* V_{tb}|^2}{|V_{cb}|^2}
\frac{6 \alpha_{QED}}{\pi f(z)} |C^{(0)eff}_{7\gamma}(\mu)|^2
\end{equation}
 where
$C^{(0)eff}_{7\gamma}(\mu)$ is the effective coefficient
for which an analytic expression can be found in \cite{BMMP:94},
 $z = {m_c}/{m_b}$, and
$f(z)$ is the phase space factor in the semileptonic
b-decay.
The expression given above is based on the
spectator model corrected for short-distance QCD effects.
Support for this approximation
comes from the $1/m_b $ expansions.
Indeed the spectator
model has been shown to correspond to the leading order approximation
in the $1/m_b$ expansion.
The next corrections appear at the ${\cal O}(1/m_b^2)$
level. The latter terms have been studied by several authors
\cite{Chay,Bj,Bigi} with the result that they affect Br(\Bsg) and
Br($B \to X_c e \bar{\nu}_e$) by only a few percent.

A critical analysis of theoretical and
experimental
uncertainties present in the prediction for Br(\Bsg) based on the
formula (\ref{main}) has been made in \cite{BMMP:94} giving
\begin{equation}
Br(B \to X_s\gamma)_{TH} = (2.8 \pm 0.8) \times 10^{-4}
\label{theo}
\end{equation}
where the error is dominated by the uncertainty in
the choice of the renormalization scale
$m_b/2<\mu<2 m_b$ as first stressed by Ali and Greub \cite{AG1} and confirmed
in \cite{BMMP:94}.
	Since \Bsg is dominated by QCD effects, it is not surprising
that this scale-uncertainty in the leading order
is particularly large.

\vspace{9.2cm}
\centerline{Fig. 5}

The \Bsg decay has already been measured and as such appears to be the only
unquestionable signal of penguin contributions! In 1993
CLEO reported \cite{CLEO}
$Br(B \to K^* \gamma) = (4.5 \pm 1.5 \pm 0.9) \times 10^{-5}.$
In 1994 the first measurement of the inclusive rate has been
presented by CLEO \cite{CLEO2}:
\begin{equation}
Br(B \to X_s\gamma) = (2.32 \pm 0.57 \pm 0.35) \times 10^{-4}
\label{incl}
\end{equation}
where the first error is statistical and the second is systematic.
This result agrees with (\ref{theo}) very well although
the theoretical and experimental errors should be decreased in
the future in order to reach a definite conclusion and to see
whether some contributions beyond the standard model
 such as present in the
Two-Higgs-Doublet Model (2HDM)
or in the Minimal Supersymmetric Standard
Model (MSSM) are required. In any case the agreement of the
theory with data is consistent with the large QCD enhancement
of \Bsg. Without this enhancement the theoretical prediction
would be at least by a factor of 2 below the data.

	Fig. 5  presents the SM prediction for the inclusive
\Bsg branching ratio including the uncertainties discussed in
\cite{BMMP:94} together with the CLEO results represented by the
shaded regions.
We stress that the theoretical result (the error bars) has been
obtained prior
to the experimental result. Since the theoretical error
is dominated by scale ambiguities  a  complete
NLO analysis is very desirable.
Such a complete next-to-leading
calculation of \Bsg is described in \cite{BMMP:94} in general terms.
As demonstrated formally there
 the cancellation of the dominant $\mu$-dependence in the leading
order can then be  achieved.

In this connection we would like to mention the analysis
of \cite{Ciu:94} in which the known two-loop mixing in the sector
$(Q_1....Q_6)$ (see table 1) has been added to the leading order
 analysis of \Bsg.
Strong renormalization scheme dependence of the resulting
branching ratio has been found, giving the branching ratio
$(1.7\pm 0.2)\cdot 10^{-4}$ and $(2.3 \pm0.3)\cdot 10^{-4}$
at $\mu=5~GeV$ for HV and NDR
schemes respectively.
This strong scheme dependence in the partial
NLO analysis presented in \cite{Ciu:94} demonstrates very
clearly the need for a full analysis.
In view of this we think
that the decrease of the branching ratio for \Bsg
relative to the LO prediction, found
in \cite{Ciu:94} and given by
$Br(B\to s\gamma)=(1.9\pm 0.2\pm 0.5)\cdot 10^{-4}$,
is still premature and one
should wait until the full NLO analysis has been done.

Finally it should be stressed that in spite of theoretical
and experimental uncertainties the CLEO measurement of
the inclusive rate provided already now an interesting
lower bound on the mass of the charged Higgs in the
most popular two Higgs doublet model. At $95\%~C.L.$
CLEO \cite{CLEO2} finds $m_{H^{\pm}}> 250~GeV$ with a similar result
obtained recently by M\"unz \cite{MM95}.
More on radiative B decays can be found
in a recent review by Ali \cite{ALI95}.
\subsection{$B\to X_s e^+e^-$ Beyond Leading Logarithms}
The rare decay $B\to X_s e^+e^-$ originates from the electroweak
penguin diagrams of fig. 1f.
The effective hamiltonian for $B\to X_s e^+e^-$ at scales $\mu=O(m_b)$
is given by
\begin{equation} \label{Heff2_at_mu}
{\cal H}_{eff}(b\to s e^+e^-) =
{\cal H}_{eff}(b\to s\gamma)  - \frac{G_F}{\sqrt{2}} V_{ts}^* V_{tb}
\left[ C_{9V}(\mu) Q_{9V}+
C_{10A}(M_W) Q_{10A}    \right]
\end{equation}
where we have again neglected the term proportional to $V_{us}^*V_{ub}$
and ${\cal H}_{eff}(b\to s\gamma)$ is given in (\ref{Heff_at_mu}).
In addition to the operators relevant for $B\to X_s\gamma$,
there are two new operators
$Q_{9V}$ and $Q_{10A}$ which are generated through the electroweak
penguin diagrams of fig. 1f and the related box diagrams needed mainly
to keep gauge invariance.
There is a large literature
on this dacay. In particular Hou et al. \cite{HWS:87} stressed
the strong dependence of $B\to X_s e^+e^-$ on $m_t$.
Further references to phenomenology can be found in \cite{BuMu:94}.
Here we concentrate on QCD corrections.

The QCD corrections to $B\to X_s e^+e^-$ have been calculated
 over the last years with increasing precision by several
groups \cite{GSW:89,GDSN:89,CRV:91,Mis:94} culminating in two complete
next-to-leading QCD calculations
\cite{Mis:94,BuMu:94} which agree with each other.
An important gain due to these NLO calculations is a considerable
reduction in the $\mu$-dependence of the resulting branching ratio.
Whereas in LO an uncertainty as large as $\pm 20\%$ can be found,
it is reduced as shown in \cite{BuMu:94} below $\pm 8\%$ after the
inclusion of NLO corrections. This is very encouraging for the
analogous efforts in $B\to X_s\gamma$ as discussed above.

An extensive numerical analysis of the differential decay rate
including NLO corrections has been presented in \cite{BuMu:94}.
As an example we show in fig. 6 the differential decay rate $R(\hat s)$
divided by $\Gamma(B \to X_c e \bar\nu)$ as
a function of $\hat s=(p_{e^+}+p_{e^-})^2/m_b^2$ for $m_t=170~GeV$ and
$\Lambda^{(5)}_{\overline{MS}}=225~MeV$. We observe that the QCD suppression
in the leading order \cite{GSW:89} is substantially weakened by the
inclusion of NLO corrections. The same result has been obtained by
Misiak \cite{Mis:94}. The $1/m^2_b$ corrections calculated in \cite{FALK}
enhance these results by roughly $10\%$.

\vspace{9.85cm}
\centerline{Fig. 6}

Finally I would like to refer to some  recent papers \cite{BSEE}, where
other aspects of $B \to X_s e^+ e^-$ have been discussed and
where further references can be found.

\subsection{ $B\to\mu\bar\mu$ and $B\to X_s\nu\bar\nu$}

$B\to\mu\bar\mu$ and $B\to X_s\nu\bar\nu$ are the theoretically
cleanest decays in the field of rare B-decays.
$B\to\mu\bar\mu$ and $B\to X_s\nu\bar\nu$
 are dominated by the $Z^0$-penguin and box diagrams
involving top quark exchanges.
Their $m_t$-dependence is fully described by the squares of
$Y(x_t)$ and $X(x_t)$ in (\ref{PBE2}) and (\ref{PBE1}) respectively.

The NLO corrections to both decays have been calculated by
Buchalla and myself \cite{BB2}. These calculations reduced
considerably the theoretical uncertainties in the branching ratios
related to the scale $\mu_t$ present in $\bar m_t(\mu_t)$.
As a by-product of this work we have also pointed out that
the previously published branching ratios $Br(B_s \to \mu\bar\mu)$
missed an overall factor of 2 which clearly has important
phenomenological implications.

I will discuss here only $B_s\to \mu\bar\mu$. Choosing
$\mu_t=m_t$ the final expression
for $B_s\to \mu\bar\mu$ including NLO corrections takes a particularly
simple form \cite{BB2}. Updating the input parameters one has \cite{BBL}:
\begin{equation}
Br(B_s\to \mu\bar\mu)=
4.18\cdot 10^{-9}\left[\frac{F_{B_s}}{230~MeV}\right]^2
\left[\frac{\bar m_t(m_t)}{170~GeV} \right]^{3.12}
\left[\frac{\mid V_{ts}\mid}{0.040} \right]^2
\left[\frac{\tau_{B_s}}{1.6 ps} \right]
\end{equation}

The impact of NLO calculations is best illustrated by giving the
theoretical uncertainty due to the choice of $\mu_t$ in the leading order
and after the inclusion of the next-to-leading corrections:

\begin{equation}
Br(B_s\to \mu\bar\mu)=(4.00\pm0.50)\cdot 10^{-9}\quad =>\quad
(4.10\pm0.05)\cdot 10^{-9}
\end{equation}
The reduction of the theoretical uncertainty is truly impressive.

Next I would like to quote the standard model expectation for
$Br(B_s\to \mu\bar\mu)$ based on the input parameters collected
in sect. 3.4 and $F_{B_s}=230\pm 40~MeV$. We find \cite{BBL}
\begin{equation}
1.7\cdot 10^{-9}\le Br(B_s\to \mu\bar\mu)\le 8.4\cdot 10^{-9}
\end{equation}
Taking on the other hand "future" input of (\ref{200}) and
$F_{B_s}=230\pm 10~MeV$ gives a smaller range:
\begin{equation}
3.1\cdot 10^{-9}\le Br(B_s\to \mu\bar\mu)\le 5.0\cdot 10^{-9}
\end{equation}

Finally I would like to stress several virtues of this very
interesting decay:
\begin{itemize}
\item
The strong dominance of the internal top exchanges and a short
distance character of the contributions allows in contrast to
$B \to X_d \gamma$ a clean determination of $\mid V_{td}\mid$
by measuring the ratio of $Br(B_d\to \mu\bar\mu)$ and
$Br(B_s\to \mu\bar\mu)$
\item
In conjunction with a future measurement of $x_s$, the branching
ratio $Br(B_s\to \mu\bar\mu)$ could help to determine
the non-perturbative parameter $B_{B_s}$ and consequently allow
a test of existing non-perturbative methods:
\begin{equation}
 B_{B_s}=
\left[\frac{x_s}{22.1}\right]
\left[\frac{\bar m_t(m_t)}{170~GeV} \right]^{1.6}
\left[\frac{4.2\cdot 10^{-9}}{Br(B_s\to \mu\bar\mu)} \right]
\end{equation}
\item
$B_s\to \mu\bar\mu$ being theoretically very clean is very
suitable for tests of physics beyond the standard model
\end{itemize}
All efforts should be made to measure this decay. The
experimental prospects are discussed in other talks in these
proceedings.

\section{Express Review of CP Violation}
\subsection{Preliminaries}
CP violation in B decays is certainly one of the most important
targets of B factories and of dedicated B experiments at hadron
facilities. It is well known that CP violating effects are expected
to occur in a large number of channels at a level attainable at
forthcoming experiments. Moreover there exist channels which
offer the determination of CKM phases essentially without any hadronic
uncertainties. Since CP violation in B decays has been already
reviewed in a special talk by Michael Gronau \cite{MG} at this meeting
and since in addition extensive reviews can be found in the literature
\cite{NQ}, let me concentrate only on the most important points.

\subsection{Strategies for $(\alpha,\beta, \gamma)$}
During the last  15 years many efforts have been made to find
means of measuring the angles in the unitarity triangle of
fig. 2 without any hadronic uncertainties. The main strategies
are as follows:

\subsubsection{$B^0$-Decays to CP Eigenstates}
A time dependent asymmetry in the decay $B^0\to f$ with $f$ being
a CP eigenstate is given by

\begin{equation}\label{e8}
a_{CP}(t,f)=
{\cal A}^{dir}_{CP}(B\to f)\cos(\Delta M
t)+{\cal A}^{mix-ind}_{CP}(B\to f)\sin(\Delta M t)
\end{equation}
where we have separated the {\it direct} CP-violating contributions
from those describing {\it mixing-induced} CP violation:
\begin{equation}\label{e9}
{\cal A}^{dir}_{CP}(B\to f)\equiv\frac{1-\left\vert\xi_f\right\vert^2}
{1+\left\vert\xi_f\right\vert^2},
\qquad
{\cal A}^{mix-ind}_{CP}(B\to f)\equiv\frac{2\mbox{Im}\xi_f}{1+
\left\vert\xi_f\right\vert^2}.
\end{equation}
In (\ref{e8}), $\Delta M$ denotes the mass splitting of the
physical $B^0$--$\bar B^0$--mixing eigenstates.
The quantity $\xi_f$ containing essentially all the information
needed to evaluate the asymmetries (\ref{e9})
is given by
\begin{equation}\label{e11}
\xi_f=\exp(-i\phi_M)\frac{A(\bar B\to f)}{A(B \to f)}
\end{equation}
with $\phi_M$ denoting the weak phase in the $B-\bar B$ mixing
and $A(B \to f)$ the decay amplitude.

Generally several decay mechanisms with different weak and
strong phases can contribute to $A(B \to f)$. These are
tree diagram (current-current) contributions, QCD penguin
contributions and electroweak penguin contributions. If they
contribute with similar strength to a given decay amplitude
the resulting CP asymmetries suffer from hadronic uncertainies
related to matrix elements of the relevant operators $Q_i$.

An interesting case arises when a single mechanism dominates the
decay amplitude or the contributing mechanisms have the same weak
phases. Then

\begin{equation}\label{e111}
\xi_f=\exp(-i\phi_M) \exp(i 2 \phi_D),
\qquad
\mid \xi_f \mid^2=1
\end{equation}
where $\phi_D$ is the weak phase in the decay amplitude.
In this particular case the hadronic matrix elements drop out,
 the direct CP violating contribution
vanishes and the mixing-induced CP asymmetry is given entirely
in terms of the weak phases $\phi_M$ and $\phi_D$. In particular
the time integrated asymmetry is given by
\begin{equation}
a_{CP}(f)=\pm \sin(2\phi_D-\phi_M)
\frac{x_{d,s}}{1+x_{d,s}^2}
\end{equation}
where $\pm$ refers to $f$ being a $CP=\pm$ eigenstate and $x_{d,s}$ are
the $B^0-\bar B^0$ mixing parameters.
Then one finds for $B_d\to \psi K_S$ and $B_d\to \pi^+\pi^-$
\begin{equation}\label{113c}
 a_{CP}(\psi K_S)=-\sin(2\beta) \frac{x_d}{1+x_d^2}
\qquad
   a_{CP}(\pi^+\pi^-)=-\sin(2\alpha) \frac{x_d}{1+x_d^2}
\end{equation}
where we have neglected for a moment QCD penguins in $a_{CP}(\pi^+\pi^-)$.
Since in the usual unitarity triangle  one side is known,
it suffices to measure
two angles to determine the triangle completely. This means that
the measurements of $\sin 2\alpha$ and $\sin 2\beta$ can determine
the parameters $\varrho$ and $\eta$.
The main virtues of this determination are as follows:
\begin{itemize}
\item No hadronic or $\Lambda_{\overline{MS}}$ uncertainties.
\item No dependence on $m_t$ and $V_{cb}$ (or $A$).
\end{itemize}

Unfortunately life is not so easy and there are only a few
channels for which this fortunate situation takes place.
In addition studies of this type require tagging ( distinction
between unmixed $B^0$ and $\bar B^0$ at $t=0$ ) as well as
two time dependent rates $B^0(t)\to f$ and  $\bar B^0(t)\to f$.
Both tagging and time dependent studies are certainly not easy.

\subsubsection{Decays to CP-Non-Eigenstates}
One can of course study also decays to CP non-eigenstates. This,
in addition to tagging, requires generally four time dependent rates
$B^0(t)\to f$,  $\bar B^0(t)\to f$, $B^0(t)\to \bar f$ and
$\bar B^0(t)\to \bar f$ \cite{DUN}. In certain cases this approach gives
interesting results.

\subsubsection{Triangle Constructions}
Here one attempts to extract $\alpha,~\beta,$ and $\gamma$ from
branching ratios only. Neither tagging nor time-dependent
measurements are needed. On the other hand these methods require
measurements of several branching ratios in order to eliminate
the hadronic uncertainties. The prototypes of such studies
are the method of Gronau and Wyler \cite{Wyler} and the
method of Gronau and London \cite{CPASYM} with the latter using the
$SU(2)$-flavour (isospin) symmetry. More recently methods
based on SU(3)-triangle relations have been proposed
\cite{grl,ghlr,hlgr}. I will return to them below.

\subsection{Theoretically Clean Determinations of
$(\alpha,\beta,\gamma)$}
In my opinion there exist only {\it five} determinations
of $(\alpha,\beta,\gamma)$ in B decays which fully deserve the
name "theoretically clean". Let me discuss them briefly one-by-one.
\subsubsection{$B_d\to\psi K_S$ and $\beta$}
The mixing induced CP-asymmetry in this "gold-plated" decay allows
 in the standard model a direct measurement of the angle $\beta$ as
 pointed out by Bigi and Sanda \cite {BSANDA} a long time ago.
In this decay the QCD penguins and EW penguins have to an
excellent approximation the same weak
phases as the leading tree contributions. This results in the
formula given in (\ref{113c}) which offers a clean determination
of $\beta$.

\subsubsection{$B_d\to\pi^+\pi^-$ and $\alpha$}
In this case the formula (\ref{113c}) deoes not really apply
because of the presence of QCD penguin contributions which have
different phases than the leading tree contributions. The
asymmetry $a_{CP}(\pi^+\pi^-)$ measures then $2 \alpha+\theta_P$
where the unknown phase $\theta_P$ signals the presence of
QCD penguins. Using the isospin symmetry and the related triangle
construction Gronau and London \cite{CPASYM}
have demonstrated how the unknown
phase $\theta_P$ can be found by measuring in addition
the branching ratios $Br(B^+\to \pi^+\pi^0)$,
$Br(B^0\to \pi^0\pi^0)$ and the branching ratios of CP conjugate
channels. With this information the asymmetry $a_{CP}(\pi^+\pi^-)$
offers a clean determination of $\alpha$.
The smallness of $Br(B_d\to \pi^0\pi^0$) ($\le{\cal O}(10^{-6})$
\cite{PAPII,KP}
is a weak point of this method.
It has been pointed out by Deshpande and He \cite{dhewp2}
that the presence of
EW penguins could have a sizable impact on the GL method.
A closer look shows, however that this impact is rather small
\cite{ghlrewp},
at most a few $\%$. Moreover as I will discuss below a
method has been proposed to estimate this effect quantitatively
\cite{PAPI,PAPIII}.

\subsubsection{$B^{\pm}\to D^0_{CP} K^{\pm}$ and $\gamma$}
This decay involves only tree diagram contributions and requires
six decay rates $B^{\pm}\to D^0_{CP} K^{\pm}$,
$B^+ \to D^0 K^+,~ \bar D^0 K^+$ and  $B^- \to D^0 K^-,~ \bar D^0 K^-$.
A known triangle construction due to Gronau and Wyler \cite{Wyler}
allows then
a clean determination of $\gamma$. The virtue of this method
is that neither tagging nor time-dependent studies are required.
Moreover the observation of CP violation in $B^{\pm}\to D^0_{CP} K^{\pm}$
would signal automatically direct CP violation. A possible difficulty is
the disparity in the size of the expected branching ratios needed
to construct the triangles in question. Whereas four branching ratios
listed above are expected to be ${\cal O}(10^{-4}-10^{-5})$, the
branching ratio of the colour supressed channels $B^{\pm}\to D^0 K^{\pm}$
are expected to be by one order of magnitude smaller. On one hand
such a small branching ratio is difficult to measure. On the other
hand the resulting triangles will have one side very small making
the extraction of $\gamma$ difficult. A similar method using
neutral B-decays has been proposed by Dunietz \cite{DUN2}.
\subsubsection{$B_s\to D^+_s K^{-}$, $\bar B_s\to D^-_s K^{+}$ and $\gamma$}
This method suggested by Aleksan, Dunietz and Kayser \cite{adk}
is unaffected
by penguin contributions. A full time dependent analysis allows a clean
measurement of $\sin^2\gamma$. Since the expected branching ratio is
${\cal O}(10^{-4})$, this method is in principle feasible although the
expected large $B^0_s-\bar B^0_s$ mixing makes this measurement very
challenging.
\subsubsection{$B_s\to\psi\phi$ and $\eta$}
This is an analog of $B_d \to \psi K_s$. In the leading order of the
Wolfenstein parametrization the asymmetry $a_{CP}(\psi\phi)$ vanishes.
Including higher order terms in $\lambda$ I find
\begin{equation}\label{DU}
a_{CP}(\psi\phi)=2\lambda^2\eta\frac{x_s}{1+x_s^2}
\end{equation}
where $\lambda$ and $\eta$ are the Wolfenstein parameters.
This agrees with Dunietz \cite{DUNS} although he expressed this
asymmetry in terms of
$\gamma$ and some poorly known CKM elements. I think the presentation
in  (\ref{DU}) is simpler: $a_{CP}(\psi\phi)$ measures $\eta$.
With $\lambda=0.22$ and $\eta=0.35$ one
has $2\lambda^2\eta=0.03$. The dilution factor suppresses the asymmetry
further and clearly the measurement of $\eta$ this way is very challenging.
Time-dependent studies at the LHC could however reach the expected level.
Moreover the very clean character of this asymmetry and its smallness
can also be used in the
search for the physics beyond the standard model.

\subsection{$SU(3)$ Triangles and Electroweak Penguins}
\subsubsection{Preliminaries}
Clearly the future B experiments will measure many more than just the
clean channels listed above and it is important to
get new ideas for other determinations of $\alpha$, $\beta$ and $\gamma$.

Last year in a series of interesting papers Gronau, Hernandez, London
and Rosner (GHLR) \cite{grl,ghlr,hlgr} used the $SU(3)$ flavour symmetry
of strong interactions
combined with certain dynamical assumptions (neglect of annihilation
diagrams etc.) to derive simple relations among B-decay amplitudes into
$\pi\pi$, $\pi K$ and $K\bar K$ final states. These $SU(3)$ relations
should allow to determine in a clean manner both the weak phases of
the CKM matrix and strong final state interaction phases by measuring
only branching ratios of the relevant B decays.

In spite of the attractiveness of this approach, it contains certain
limitations. Let me first mention the one discussed in \cite{bf}.
Irrespectively of the uncertainties related to SU(3)-breaking,
discussed in \cite{ghlrsu3},
the success of this approach depends in the case of $\beta$ on
whether the QCD penguin amplitudes are fully dominated by the
diagrams with internal top quark exchanges. As we have shown in
\cite{bf} this full dominance cannot be justified and the presence of
QCD penguin contributions with internal u- and c- quarks
precludes a clean determination of the angle $\beta$ by using
this approach. On the other hand, as we have shown, the
determination of the angle $\gamma$ is unaffected by these new
contributions.

Yet there are still other contributions which were not taken
into account by GHLR. These are the electroweak penguin contributions.
As pointed out by Deshpande and He \cite{dhewp2} the EW penguins do have an
impact on the GHLR approach even in the case of $\gamma$.
Before discussing this let me briefly review the importance of EW penguins
in B decays in general.
\subsubsection{Electroweak Penguins and B decays}
During the last two years there has been a considerable interest in
the role of electroweak penguin contributions in non-leptonic $B$-decays.
Since the Wilson coefficients of the corresponding local operators
increase strongly with the top-quark mass, it has been found by
Robert Fleischer \cite{rfewp1,rfewp2,rfewp3} that the role of the electroweak
penguins can be substantial in certain decays. This is for instance
the case of the decay $B^-\to K^-\Phi$ \cite{rfewp1},
which  exhibits sizable electroweak penguin effects.
More interestingly, there are even some channels, such as $B^-\to\pi^-\Phi$
\cite{rfewp2} and $B_s\to\pi^0\Phi$ \cite{rfewp3}, which are
{\it dominated completely} by electroweak penguin contributions
and which should, thus, allow interesting insights into the physics
of the corresponding operators. In this respect, the
decay $B_s\to\pi^0\Phi$ (or similar transitions such as $B_s\to\rho^0\Phi$)
is very promising due to its special isospin-, CKM- and
colour-structure \cite{rfewp3}. As the branching ratio of this mode
is expected to be of ${\cal O}(10^{-7})$, it will unfortunately be
rather difficult to analyze this decay experimentally.
The electroweak penguin effects discussed in \cite{rfewp1,rfewp3}
have been confirmed by other authors \cite{dhewp}-\cite{dy}.

In the foreseeable future the branching ratios of ${\cal O}(10^{-5})$
and possibly ${\cal O}(10^{-6})$ will be experimentally available and
it is important to ask about the role of electroweak penguin effects
in the corresponding channels. In particular, the question arises whether
the usual strategies for the determination of the CKM-phases are
affected by the presence of the electroweak penguin contributions.

As we have stated above the five methods reviewed in section 6.3 are,
 -- except for the Gronau-London method --
unaffected by EW penguins. It can
also be shown that the impact of the $\bar b \to \bar d$ EW penguins
on the Gronau-London method is small \cite{ghlrewp,PAPI}.
On the other hand the point of Despande and He \cite{dhewp2}
that
the $\bar b \to \bar s$ EW penguins have a considerable impact on
the GHLR method is now well accepted.

In this connection GHLR have presented  a
systematic classification of electroweak penguins in two-body
$B$-decays \cite{ghlrewp}. Moreover,
in this paper they have constructed an amplitude quadrangle
for $B\to\pi K$ decays that can be used -- at least in principle --
to extract the CKM-angle $\gamma$ irrespectively of the presence of
electroweak penguins. Unfortunately, from the experimental point of view
this approach is rather difficult, because one diagonal of the
quadrangle corresponds to the amplitude of the electroweak penguin
dominated $B_s$-decay $B_s\to\pi^0\eta$ which is expected to have
a very small branching ratio at the ${\cal O}(10^{-7})$ level. Another
$SU(3)$-symmetry based method of extracting $\gamma$, where electroweak
penguins are also eliminated, has been presented by
Deshpande and He~\cite{dhgam}. Although this approach using the
charged $B$-decays $B^-\to\{\pi^-\bar K^0,\pi^0K^-,\eta K^-\}$ and
$B^-\to\pi^-\pi^0$ should be more promising for experimentalists,
it is affected by $\eta$--$\eta'$--mixing and other $SU(3)$-breaking
effects and therefore cannot be regarded as a clean measurement
of $\gamma$.
\subsubsection{Towards the Control over Electroweak Penguins}
In view of this situation, it would be useful
to determine the electroweak penguin contributions experimentally.
Once this has been achieved, their role in a variety of $B$-decays
could be explicitly found. Although some thoughts on this issue have
appeared in~\cite{ghlrewp}, no constructive quantitative
method has been proposed there.

Here I would like to report on a work of Fleischer and myself \cite{PAPI}
in which we have suggested a
different ``philosophy'' of applying the $SU(3)$ amplitude relations.
In contrast to GHLR, we think that these relations are
more useful from the phenomenological point of view if one uses
the phase $\gamma$ {\it as one of the central inputs}.
As we have stated above,
there are already methods on the market allowing a measurement
of this phase in an {\it absolutely clean way} without any effect
coming from the electroweak penguins.

At first sight, this new philosophy might appear not useful because
one of the goals of the GHLR strategy was precisely the
determination of $\gamma$. Yet, as we have seen, this program
is difficult to realize without further inputs. On the other
hand, as shown in \cite{PAPI}, once the phase $\gamma$ is used as an
input, the electroweak penguin contributions can be straightforwardly
determined. This knowledge subsequently allows the determination
of CKM-phases in a variety of $B$-decays \cite{PAPI,PAPIII}. Consequently,
with this new strategy, the GHLR method is resurrected. Moreover, the
impact of electroweak penguins on the $\alpha$-determination using
$B(\bar B)\to\pi\pi$ decays can be \mbox{{\it quantitatively}} estimated.

Our strategy proceeds in three steps. In the first step $\gamma$ is
taken form a decay without EW-penguin contributions (see for instance
sections 6.3.3 or 6.3.4). In the second step one measures the branching
ratios for $B^+\to\pi^+ K^0$, $B^+\to\pi^0 K^+$, $B^-\to\pi^0 K^-$,
$B_d^0\to\pi^-K^+$ and $\bar B_d^0\to \pi^+ K^-$ and using only
$SU(2)$ symmetry finds the $\bar b \to \bar s$ EW-penguin.
In the third step one takes $\beta$ from a clean decay like
$B\to \psi K_S$ and using $SU(3)$ symmetry determines the small
$\bar b \to \bar d$ EW-penguin.
Note that in this method only branching ratios ${\cal O}(10^{-5})$
are required.

Having determined $\bar b \to \bar s$ and $\bar b \to \bar d$ EW
penguins this way one can study their effects in other decays.
In particular their effect on the $\alpha$-determination by
means of $B\to\pi\pi$ can be studied quantitatively.
More details on this method and an alternative strategy can
be found in \cite{PAPI}.
\subsubsection{Other Ideas}
The Gronau-London method for the $\alpha$-determination involves the
experimentally difficult mode $B_d \to \pi^0\pi^0$ which is
expected to be $\le {\cal O}(10^{-6})$ \cite{KP}.
Moreover it is slightly affected by EW penguin contributions.
In \cite{PAPII} we have presented an alternative method of
determining $\alpha$ by performing simultaneous measurements
of the mixing-induced CP asymmetries of the decays
$B_d\to\pi^+\pi^-$ and $B_d\to K^0\bar K^0$. The accuracy of
this method is limited by $SU(3)$ breaking effects but it is
unaffected by EW penguins. Moreover the decay $B_d \to K^0\bar K^0$
might be easier to measure than $B_d \to \pi^0\pi^0$. In this
connection I would like to mention that in contrast to previous
claims the CP asymmetry in $B_d \to K^0\bar K^0$ is non-vanishing
in the standard model. Indeed as shown by Fleischer \cite{PAPF}
the inclusion of QCD penguins with internal u- and c-quarks in
$B_d \to K^0\bar K^0$ can result in a CP asymmetry as large as
${\cal O}(50\%)$.

Next I would like to mention a recent interesting paper by Fleischer
\cite{PAPIII} where various strategies for fixing the angle $\gamma$
and obtaining experimental insight into the world of EW-penguins have been
presented. A by-product of this work is a refined estimate of the role
of EW-penguins in the $\alpha$-determination by means of $B\to\pi\pi$
decays.

\section{Classification}
It is probably a good place to group the various decays and
quantities into three distinct classes. I include in this
classification also K-decays.
\subsection{Class 1}
These are the decays with essentially no theoretical uncertainties:
\begin{itemize}
\item
CP violation in $B_d\to \psi K_S$, $B^{\pm}\to D_{CP}K^{\pm}$,
 $B_s\to D_sK$
$B_s\to\psi\phi$, $B\to \pi\pi$,
\item
Rare B-decays: $B_d\to l\bar l$, $B_d\to X_s\nu\bar\nu$, $x_d/x_s$
\item
Rare K-decays: $K_L \to \pi^0\nu\bar\nu$, $K^+\to \pi^+\nu\bar\nu$
\end{itemize}
\subsection{Class 2}
Here I group the quantities or decays with moderate theoretical
uncertainties which should be considerably reduced in the next
five years. In particular I assume that the uncertainties in $B_K$
and $F_B\sqrt{B}$ will be reduced below 10\% and that the NLO
corrections to $B\to X_s \gamma$ will be completed.
\begin{itemize}
\item
$B\to X_s\gamma$, $B\to X_s e^+ e^-$, $B\to K^*e^+e^-$
\item
$x_d,~ x_s$, $\mid V_{cb}\mid_{excl}$, $\mid V_{cb}\mid_{incl}$,
$\mid V_{ub}/V_{cb}\mid_{incl}$
\item
Some CP asymmetries in B-decays discussed in section 6.
\item
$\varepsilon_K$ and $K_L\to \pi^0 e^+e^-$
\end{itemize}
\subsection{Class 3}
Here we have a list of important decays with large theoretical
uncertainties which can only be removed by a dramatic progress
in non-perturbative techniques:
\begin{itemize}
\item
CP asymmetries in most $B^{\pm}$-decays
\item
$B_d\to K^*\gamma$, Non-leptonic B-decays, $\mid V_{ub}/V_{cb}\mid_{excl}$
\item
$\varepsilon^{\prime} /\varepsilon$, $K\to \pi\pi$, $\Delta M(K_L-K_s)$,
$K_L\to\mu\bar\mu$, hyperon decays and so on.
\end{itemize}
It should be stressed that even in the presence of theoretical
uncertainties a measurement of a non-vanishing
ratio $\varepsilon^{\prime}/\varepsilon$ or a non-vanishing CP asymmetry
in charged B-decays would signal direct CP violation excluding
superweak scenarios \cite{WO1}. This is not guaranteed by
several clean decays of class 1 \cite{WIN} except for
$B^{\pm}\to D_{CP} K^{\pm}$.

\section{Future Visions}
Let us next concentrate on decays of class 1 which as we have stated
above are essentially free of hadronic uncertainties.
As we have seen in section 3 using the standard analysis  with
rather optimistic assumptions
about the theoretical and experimental errors it was  difficult to
achieve the accuracy better than $\Delta\varrho=\pm 0.10$ and
$\Delta\eta=\pm 0.05$. Let us then see what can be achieved
with the decays of class 1.

\subsection{$(\rho,\eta)$ from $(\sin(2\alpha),\sin(2\beta))$}
With $a\equiv \sin(2\alpha)$ and $b\equiv \sin(2\beta)$
one determines $\varrho$, $\eta$
as follows \cite{AB94}:
\begin{equation}\label{5}
\bar\varrho = 1-\bar\eta r_{+}(b)\quad ,\quad
\bar\eta=\frac{r_{-}(a)+r_{+}(b)}{1+r_{+}^2(b)}
\end{equation}
where
\begin{equation}\label{7}
r_{\pm}(z)=\frac{1}{z}(1\pm\sqrt{1-z^2})
\qquad
z=a,b
\end{equation}
where $(\bar\varrho,\eta)$ are defined in (\ref{3}).
As illustrative examples we consider in table 4 three scenarios.
The first two rows give the assumed input parameters and their
experimental errors. The remaining rows give the results for
$\varrho$, $\eta$ and $\mid V_{ub}/V_{cb}\mid$.
Further results can be found in \cite{AB94}.
The accuracy in the scenario I should be achieved at B-factories
and HERA-B.
 Scenarios II and
III correspond potentially to B-physics at Fermilab during
the Main Injector era and at LHC respectively.

\begin{table}
\begin{center}
\begin{tabular}{|c||c||c|c|c|}\hline
& Central &$I$&$II$&$III$\\ \hline
$\sin(2\alpha)$ & $0.40$ &$\pm 0.08$ &$\pm 0.04$ & $\pm 0.02 $\\ \hline
$\sin(2\beta)$ & $0.70$ &$\pm 0.06$ &$\pm 0.02$ & $\pm 0.01 $\\ \hline
\hline
$\varrho$ &$0.072$ &$\pm 0.040$&$\pm 0.016$ &$\pm 0.008$\\ \hline
$\eta$ &$0.389$ &$\pm 0.044$ &$\pm 0.016$&$\pm 0.008$ \\ \hline
$\mid V_{ub}/V_{cb}\mid$ &$0.087$ &$\pm 0.010$ &$\pm 0.003$&$\pm 0.002$
 \\ \hline
\end{tabular}
\end{center}
\centerline{}
\caption{Determinations of $\varrho$ and $\eta$ using $\sin(2\alpha)$
and $\sin(2\beta)$ }
\label{TAB4}
\end{table}
Table 4 shows very clearly the potential of CP asymmetries
in B-decays in the determination of CKM parameters.
It should be stressed that this high accuracy is not only achieved
because of our assumptions about future experimental errors in the
scenarios considered, but also because $\sin(2\alpha)$ is a
very sensitive function of $\varrho$ and $\eta$ \cite{BLO},
and most importantly because
of the clean character of the quantities considered.
\subsection{$(\rho,\eta)$ from $(R_t,\sin(2\beta))$}
An alternative strategy is to use the measured value of $R_t$ instead
of $\sin(2\alpha)$. A clean measurement of $R_t$ can be achieved
using the ratio $x_d/x_s$. Then (\ref{5})
is replaced by \cite{AB94A}
\begin{equation}\label{5a}
\bar\varrho = 1-\bar\eta r_{+}(b)\quad ,\quad
\bar\eta=\frac{R_t}{\sqrt{2}}\sqrt{b r_{-}(b)}
\end{equation}
The result of this exercise is shown in table 5.
Although this determination of
CKM parameters cannot fully compete with the previous one the
consistency of both determinations will offer an important test of
the standard model.

\begin{table}
\begin{center}
\begin{tabular}{|c||c||c|c|c|}\hline
& Central &$I$&$II$&$III$\\ \hline
$R_t$ & $1.00$ &$\pm 0.10$ &$\pm 0.05$ & $\pm 0.03 $\\ \hline
$\sin(2\beta)$ & $0.70$ &$\pm 0.06$ &$\pm 0.02$ & $\pm 0.01 $\\ \hline
\hline
$\varrho$ &$0.076$ &$\pm 0.111$&$\pm 0.053$ &$\pm 0.031$\\ \hline
$\eta$ &$0.388$ &$\pm 0.079$ &$\pm 0.033$&$\pm 0.019$ \\ \hline
$\mid V_{ub}/V_{cb}\mid$ &$0.087$ &$\pm 0.014$ &$\pm 0.005$&$\pm 0.003$
 \\ \hline
\end{tabular}
\end{center}
\centerline{}
\caption{Determinations of $\varrho$ and $\eta$ using $R_t$
and $\sin(2\beta)$.}
\label{TAB5}
\end{table}

\subsection{$\sin(2\beta)$ from $K\to \pi\nu\bar\nu$}
It has been pointed out in \cite{BB4} that
measurements of $Br(K^+\to \pi^+\nu\bar\nu)$ and
$Br(K_L\to \pi^0\nu\bar\nu)$ offer
an interesting measurement of $\sin(2\beta)$ without hadronic
uncertainties.
Choosing
$Br(K^+\to \pi^+\nu\bar\nu)=(1.0\pm 0.1)\cdot 10^{-10}$ and
$Br(K_L\to \pi^0\nu\bar\nu)=(2.5\pm 0.25)\cdot 10^{-11}$,
one finds \cite{BB4}
\begin{equation}\label{26}
\sin(2 \beta)=0.60\pm 0.06 \pm 0.03 \pm 0.02
\end{equation}
where the first error is "experimental", the second represents the
uncertainty in $m_c$ and  $\Lambda_{\overline{MS}}$ and the last
is due to the residual renormalization scale uncertainties.
The $m_t$ and $V_{cb}$ dependences are negligible. This
determination of $\sin(2\beta)$ is competitive with the one
expected at the B-factories at the beginning of the next decade.

Other future visions can be found in \cite{BLO,AB94,AB94A}.

\section{Summary}
In this review we have discussed the most interesting quantities
in B physics which
when measured should have an
important impact on our understanding of CP
violation and  quark mixing.
We have also stressed that K physics can also offer an important
contribution to this issue.

In this  review we have concentrated on rare decays and  CP violation in the
standard model. The structure of rare decays and of CP violation in
extensions of the
standard model may deviate from this picture.
Consequently the situation in this field could turn out to be very different
from the one presented here.
However in order to distinguish the standard model predictions from
the predictions of its extensions it is essential that the
theoretical calculations reach acceptable precision. In this
context we have emphasized the importance of the QCD calculations in
rare and CP violating decays. During the recent years a considerable
progress has been made in this field through the computation of NLO
contributions to a large class of decays \cite{BBL}.
 This effort reduced considerably
the theoretical uncertainties in the relevant formulae and thereby improved
the determination of the CKM parameters to be achieved in future
experiments. At the same time it should be stressed that whereas the
theoretical status of QCD calculations for rare semileptonic decays like
$K \to \pi\nu\bar\nu$, $B\to \mu\bar\mu$, $B \to X_s e^+ e^-$
 is fully satisfactory and
the status of $B\to X_s\gamma$ should improve in the coming years, a lot
remains to be done in a large class of non-leptonic decays or transitions
where non-perturbative uncertainties remain sizable.

Clearly an exciting future is ahead of us. Let us just imagine that
the following quantities have been measured to an acceptable accuracy:
\begin{itemize}
\item
$\left| V_{ub}/V_{cb}\right|$, $\left| V_{cb}\right|$, $m_t$
\item
$Br(K^+\to\pi^+\nu\bar\nu)$, $Br(K_L\to\pi^0\nu\bar\nu)$,
$Br(B_s\to\mu^+\mu^-)$,
\item
$B^0_s-\bar B^0_s$- mixing
\item
$B\to X_s\gamma$ and $B \to X_s e^+ e^-$
\item
$(\alpha,\beta,\gamma)$ in CP asymmetries in B decays.
\item
$\varepsilon^{\prime}/\varepsilon$ and $K_L\to\pi^0 e^+ e^-$.
\end{itemize}
and various non-perturbative parameters like $B_K$, $F_B\sqrt{B}$,
and $(B_6,B_8)$ in  $\varepsilon^{\prime}/\varepsilon$ have been
calculated to
an acceptable precision. With all these things at our disposal we
could really get a great insight into the physics of CP violation
and quark mixing and in particular find possible signals of
some new physics beyond the standard model. At present all this
is only a dream but already in ten years, at "Beauty 05", it could
turn out to be reality!

\section{Final Remarks}
"Beauty 95" will remain  in my memory for a long time and I am sure that
this will also be the case for all participants. It was a splendid meeting
run by the first lady Sue Geddes and the three musketeers Roger
Cashmore, Neville Harnew and Peter Schlein helped by the team
consisting of David Bailey, Lino Demaria, Anish Grewal, Chris Parkes,
Armin Reichold and Davide Vite. They created five most enjoyable
days in the scenery of Wadham College. Many thanks to them!

Yet nothing can be perfect. There was something missing at this meeting.
I am referring here to four missing lambs at the Maytime Inn, which in
the case of Fernando, Michael, Simone and myself, have been replaced
by  four dull chickens. Let us hope
that during "Beauty 00" or "Beauty 05", held hopefully in Oxford again,
the Maytime Inn will either find the missing lambs or serve us penguins
instead.

\vfill\eject


\begin{thebibliography}{999}
\bibitem{GF}
{\sc R.P. Feynman and M. Gell-Mann,}
 {\sl Phys. Rev.} {\bf 109} (1958) 193.
\bibitem{BJLW}
{\sc A.J. Buras, M.Jamin and M.E. Lautenbacher,}
{\sl Nucl. Phys.} {\bf B 408} (1993) 209.
\bibitem{NEU}
{\sc M. Neubert}
{\sl Physics Reports}, {\bf 245} (1994) 259.
\bibitem{PBE}
{\sc G. Buchalla, A.J. Buras and M.K. Harlander,} {\sl Nucl. Phys.}
 {\bf B 349} (1991) 1.
\bibitem{BSS}
{\sc A.J. Buras, W. Slominski and H. Steger,}
{\sl Nucl. Phys.} {\bf B 238} (1984) 529;
{\sl Nucl. Phys.} {\bf B 245} (1984) 269.
\bibitem{IL}
{\sc T. Inami and C.S. Lim,}
{\sl Progr. Theor. Phys.} {\bf 65} (1981) 297.
\bibitem{Chay}
{\sc J.~Chay, H.~Georgi and B.~Grinstein,}
{\sl Phys.~Lett.} {\bf B247} (1990) 399.
\bibitem{Bj}
{\sc J.~D.~Bjorken, I.~Dunietz and J.~Taron,}
{\sl Nucl.~Phys.} {\bf B371} (1992) 111.
\bibitem{Bigi} {\sc I.~I.~Bigi, B.~Blok, M.~Shifman, N.~G.~Uraltsev and
		A.~I.~Vainshtein,} in "B-Decays" (2nd Edition) edited by
            S. Stone, World Scientific (1994) page 132.
		{\sc I.I. Bigi} {\sl et al,}
{\sl Phys. Lett.} {\bf B 293} (1992) 430; Erratum {\bf 297} (1993) 477;
{\sl Phys. Lett.} {\bf B 323} (1994) 408;
{\sl Phys. Rev. Lett.} {\bf 71} (1993) 496;
\bibitem{Mannel}
{\sc A.V. Manohar and M.B. Wise}
{\sl Phys. Rev.} {\bf D 49} (1994) 1310;
{\sc A.F. Falk, M.Luke, and M.J. Savage,}
{\sl Phys. Rev.} {\bf D 49} (1994) 3367;\\
{\sc T. Mannel,}{\sl Nucl.~Phys.} {\bf B413} (1994) 396;
{\sc M. Neubert}, {\sl Phys. Rev.} {\bf D 49} (1994) 3392 and 4623;
\bibitem{Beneke}
{\sc I.I. Bigi et al.} {\sl Phys. Rev.} {\bf D 50} (1994) 2234;
{\sc M. Beneke, V.M. Braun and V.I. Zakharov,}
{\sl Phys. Rev. Lett.} {\bf 73} (1994) 3058;
{\sc M. Beneke and V.M. Braun,} {\sl Nucl.~Phys.} {\bf B426} (1994) 301.
\bibitem{Braun}
{\sc P. Ball, M. Beneke and V.M. Braun,} CERN-TH/95-65, hep-ph/9503492.
\bibitem{ALTA}
{\sc G. Altarelli, G. Curci, G. Martinelli and S. Petrarca,}
{\sl Nucl. Phys.} {\bf B 187} (1981) 461.
\bibitem{BW}
{\sc A.J. Buras and P.H. Weisz,}
{\sl Nucl. Phys.} {\bf B 333} (1990) 66.
\bibitem{BBL}
{\sc G. Buchalla, A.J. Buras and M. Lautenbacher,} in preparation.
 \bibitem{BJLW1}
{\sc A.J. Buras, M.Jamin, M.E. Lautenbacher and P.H. Weisz,}
{\sl Nucl. Phys.} {\bf B 370} (1992) 69;
{\sl Nucl. Phys.} {\bf B 400} (1993) 37.
\bibitem{BJLW2}
{\sc A.J. Buras, M.Jamin and M.E. Lautenbacher,}
{\sl Nucl. Phys.} {\bf B 400} (1993) 75.
\bibitem{ROMA1}
{\sc M. Ciuchini, E. Franco, G. Martinelli and L. Reina,}
{\sl Phys. Lett.} {\bf B 301} (1993) 263.
\bibitem{ROMA2}
{\sc M. Ciuchini, E. Franco, G. Martinelli and L. Reina,}
{\sl Nucl. Phys.} {\bf B 415} (1994) 403.
\bibitem{MisMu:94}
{\sc M.Misiak and M. M{\"u}nz,}
{\sl Phys. Lett.} {\bf B344} (1995) 308.
\bibitem{Buch:93}
{\sc G. Buchalla,} {\sl Nucl. Phys.} {\bf B 391} (1993) 501.
\bibitem{Bagan}
{\sc E. Bagan, P.Ball, V.M. Braun and P.Gosdzinsky,}
{\sl Nucl. Phys.} {\bf B 432} (1994) 3;
{\sc E. Bagan} {\sl et al.,} {\sl Phys. Lett.} {\bf B 342} (1995) 362;
{\bf B 351} (1995) 546;
\bibitem{JP}
{\sc M. Jamin and A. Pich,}
{\sl Nucl.~Phys.} {\bf B425} (1994) 15.
\bibitem{HNa}
{\sc S. Herrlich and U. Nierste,}
{\sl Nucl. Phys.} {\bf B419} (1994) 292.
\bibitem{BJW}
{\sc A.J. Buras, M. Jamin, and P.H. Weisz,}
{\sl Nucl. Phys.} {\bf B 347} (1990) 491.
\bibitem{HNb}
{\sc S. Herrlich and U. Nierste}
{\bf TUM-T31-81/95; hep-ph 9507262}
\bibitem{BB1}
{\sc G. Buchalla and A.J. Buras,}
{\sl Nucl. Phys.} {\bf B 398} (1993) 285.
\bibitem{BB2}
{\sc G. Buchalla and A.J. Buras,}
{\sl Nucl. Phys.} {\bf B 400} (1993) 225.
\bibitem{BB3}
{\sc G. Buchalla and A.J. Buras,}
{\sl Nucl. Phys.} {\bf B 412} (1994) 106.
\bibitem{BB5}
{\sc G. Buchalla and A.J. Buras,}
{\sl Phys. Lett.} {\bf B 336} (1994) 263.
\bibitem{BLMM}
{\sc A. J. Buras, M. E. Lautenbacher, M. Misiak and M. M{\"u}nz,}
{\sl Nucl.~Phys.} {\bf B423} (1994) 349.
\bibitem{Mis:94}
{\sc M. Misiak,}
{\sl Nucl.~Phys.} {\bf B393} (1993) 23;
{\sl Erratum}, {\sl Nucl.~Phys.} {\bf B439} (1995) 461.
\bibitem{BuMu:94}
{\sc A.J. Buras and M. M{\"u}nz,}
{\sl Phys. Rev.} {\bf D 52} (1995) 186.
\bibitem{CAB}
{\sc {N. Cabibbo}}, {\sl Phys. Rev. Lett.} {\bf 10} (1963) 531.
\bibitem{KM}
{\sc M. Kobayashi and K. Maskawa},
 {\sl Prog. Theor. Phys.} {\bf 49} (1973) 652.
\bibitem{WO}
{\sc L. Wolfenstein}, {\sl Phys. Rev. Lett.} {\bf 51} (1983) 1945.
\bibitem{BLO}
{\sc A.J. Buras, M.E. Lautenbacher and G. Ostermaier,}
{\sl Phys. Rev.} {\bf D 50} (1994) 3433.
\bibitem{PDG}
{\sc Particle Data Group,} {\sl Phys. Rev.} {\bf D 50} (1994) 1.
\bibitem{CDF}
{\sc F. Abe }  {\sl et al.}, {\sc CDF},
 {\sl Phys. Rev.} {\bf D 50} (1994) 2966,
 {\sl Phys. Rev. Lett.} {\bf 73} (1994) 225,
{\sl Phys. Rev.} {\bf D 51} (1995) 4623.
\bibitem{D0}
{\sc S. Abachi}  {\sl et al.},{\sc D0}, FERMILAB-PUB-95/028E (1995).
\bibitem{PAT}
{\sc R. Patterson}, in Proc. of the XXVII Int. Conf. on High Energy Physics,
(Glasgow 94), edited by P.J. BUSSEY and I.G. KNOWLES (IOP Publications Ltd.,
Bristol,1995), Vol,p.149.
\bibitem{Neubert}
{\sc M. Neubert,} {\sl Phys. Lett.} {\bf B 338} (1994) 84.
\bibitem{SUV}
{\sc M. Shifman, N.G. Uraltsev and A. Vainshtein,}
{\sl Phys. Rev.} {\bf D 51} (1995) 2217.
\bibitem{URAL}
{\sc N.G. Uraltsev}, TPI-MINN-95/5-T.
\bibitem{CLEO3}
{\sc E. Thorndike} (CLEO), EPS-HEP Conference, Brussels, July, 1995.
\bibitem{AB}
{\sc A.J. Buras,} {\sl Phys. Lett.} {\bf B 317} (1993) 449.
\bibitem{SH0}
{\sc S. Sharpe,}
{\sl Nucl. Phys. (Proc. Suppl.)} {\bf B34} (1994) 403;
\bibitem{Ishizuka}
{\sc N. Ishizuka} et al.,
{\sl Phys. Rev. Lett.} {\bf 71} (1993) 24.
\bibitem{BBG0}
{\sc W.A. Bardeen, A.J. Buras and J.-M. G\'erard,}
{\sl Phys. Lett.} {\bf B211} (1988) 343;
 {\sc J-M. G\'erard,} {\sl Acta Physica Polonica} {\bf B21} (1990) 257.
\bibitem{Bijnens}
{\sc J. Bijnens and J. Prades,} {\sl Nucl. Phys.} {\bf B 444} (1995) 523.
\bibitem{Prades}
{\sc A. Pich and E. de Rafael,} {\sl Phys. Lett.} {\bf B158} (1985) 477;
{\sc J. Prades} {\sl  et al,} {\sl Z. Phys.}  {\bf C51} (1991) 287.
\bibitem{Donoghue}
{\sc J.F. Donoghue, E. Golowich and B.R. Holstein,}
{\sl Phys. Lett.} {\bf B119} (1982) 412.
\bibitem{Chris}
{\sc C.T. Sachrajda,}
 in "B-Decays" (2nd Edition) edited by
            S. Stone, World Scientific (1994) page 602.
\bibitem{Duncan}
{\sc A. Duncan, E. Eichten, J. Flynn, B. Hill, G. Hockney and H. Thacker,}
{\sl Phys. Rev.} {\bf D 51} (1995) 5101.
\bibitem{Latt}
{\sc C.W. Bernard, J.N. Labrenz and A. Soni,}
{\sl Phys. Rev.} {\bf D 49} (1994) 2536; {\sc T. Draper and C. McNeile,}
{\sl Nucl. Phys. (Proc. Suppl.)} {\bf 34} (1994) 453.
\bibitem{QCDS}
{\sc E. Bagan, P. Ball, V.M. Braun and H.G. Dosch},
{\sl Phys. Lett.} {\bf B 278} (1992) 457;
{\sc M. Neubert}, {\sl Phys. Rev.} {\bf D 45} (1992) 2451;
\bibitem{BGL}
{\sc C.G. Boyd, B. Grinstein and R.F. Lebed}, UCSD/PTH 94-27.
\bibitem{NAR}
{\sc S. Narison,}
{\sl Phys. Lett.} {\bf B322} (1994) 247;
\bibitem{Rizzo}
{\sc G. Rizzo, these proceedings.}
\bibitem{Stocchi}
{\sc A. Stocchi, these proceedings.}
\bibitem{Aleksan}
\sc{R. Aleksan,} Plenary talk, EPS-HEP Conference, Brussels, July, 1995.
\bibitem{RUT}
{\sc M. Ciuchini et al.}
{\bf CERN-TH-7514-94, hep-ph/9501265}
\bibitem{ALUT}
{\sc A. Ali and D. London,}
{\bf DESY 95-148, hep-ph 9508272} (1995).
\bibitem{PW}
{\sc R.D. Peccei and K. Wang,}
{\sl Phys. Lett.} {\bf B 349} (1995) 220.
\bibitem{AP:92}
{\sc G. Altarelli and S. Petrarca,}
{\sl Phys. Lett.} {\bf B 261} (1991) 303.
\bibitem{Browder}
{\sc T.E. Browder and K. Honscheid,} UH-511-816-95,
{\bf hep-ph/9501287}, to be published in Progress in Particle and Nuclear
Physics {\bf 35}.
\bibitem{Pham}
{\sc Q. Hokim and X.Y. Pham,} {\sl Phys. Lett.} {\bf B 122} (1983) 297;
{\sl Ann. Phys.} {\bf B 155} (1984) 202.
\bibitem{Voloshin}
{\sc M.B. Voloshin,}
{\sl Phys. Rev.} {\bf D 51} (1995) 3948;
\bibitem{nc}
{\sc I.I. Bigi et al.,} {\sl Phys. Lett.} {\bf B 323} (1994) 408;
{\sc A.F. Falk, M.B. Wise and I. Dunietz},
{\sl Phys. Rev.} {\bf D 51} (1995) 1183; {\sc I. Dunietz,} (hep-ph/9501287).
\bibitem{BDY}
{\sc G. Buchalla, I. Dunietz and H. Yamamoto},
FERMILAB-PUB-95/167-T.
\bibitem{RR}
{\sc R. R\"uckl,} these proceedings.
\bibitem{STECH}
{\sc M. Wirbel, B. Stech and M. Bauer},
{\sl Z.Phys.} {\bf C29} (1985) 637.
\bibitem{AB:95c}
{\sc A.J. Buras,}
{\sl Nucl. Phys.} {\bf B 434} (1995) 606.
\bibitem{Bjj}
{\sc J.D. Bjorken,} {\sl Nucl. Phys. (Proc. Suppl.) }
{\bf B 434} (1989) 325; SLAC-PUB-5389.
\bibitem{Bert}
{\sc S.~Bertolini, F.~Borzumati and A.~Masiero,}
{\sl Phys. Rev. Lett.} {\bf 59} (1987) 180.
\bibitem{Desh}
{\sc N.~G.~Deshpande, P.~Lo, J.~Trampetic, G.~Eilam and P. Singer}
{\sl Phys. Rev. Lett.} {\bf 59} (1987) 183.
\bibitem{Grin}
{\sc B.~Grinstein, R.~Springer and M.B.~Wise,}
{\sl Nucl.~Phys.} {\bf B339} (1990) 269.
\bibitem{Odon}
{\sc R.~Grigjanis, P.J.~O'Donnell, M.~Sutherland and H.~Navelet,}
{\sl Phys.~Lett.} {\bf B213} (1988) 355;
{\sl Phys.~Lett.} {\bf B286} (1992) 413 E.
\bibitem{CFMRS:93}
{\sc M. Ciuchini, E. Franco, G. Martinelli, L. Reina and L. Silvestrini,}
 {\sl Phys.~Lett.} {\bf B316} (1993) 127.
\bibitem{CFRS:94}
{\sc M. Ciuchini, E. Franco, L. Reina and L. Silvestrini,}
{\sl Nucl.~Phys.} {\bf B421} (1994) 41.
\bibitem{CCRV:94a}
{\sc G.~Cella, G.~Curci, G.~Ricciardi and  A.~Vicer{\'e},}
{\sl Phys.~Lett.} {\bf B325} (1994) 227.
\bibitem{CCRV:94b}
{\sc G.~Cella, G.~Curci, G.~Ricciardi and A.~Vicer{\'e},}
{\sl Nucl.~Phys.} {\bf B431} (1994) 417.
\bibitem{Yao1} {\sc K.~Adel and Y.P.~Yao,}
{\sl Modern Physics Letters} {\bf A8} (1993) 1679;
{\sl Phys. Rev.} {\bf D 49} (1994) 4945.
\bibitem{AG1}
{\sc A.~Ali, and  C.~Greub,} {\sl Z.Phys.} {\bf C60} (1993) 433.
\bibitem{AG2}
{\sc A.~Ali, and  C.~Greub,} {\sl Z.Phys.} {\bf C49} (1991) 431,
{\sl Phys.~Lett.} {\bf B259} (1991) 182.
\bibitem{BMMP:94}
{\sc A. J. Buras, M. Misiak, M. M{\"u}nz and S. Pokorski,}
{\sl Nucl.~Phys.} {\bf B424} (1994) 374.
\bibitem{CLEO}
{\sc R.~Ammar} {\sl et. al.} (CLEO),
{\sl Phys. Rev. Lett.} {\bf 71} (1993) 674.
\bibitem{CLEO2} {\sc M.S. Alam} {\sl et. al} (CLEO),
{\sl Phys. Rev. Lett.} {\bf 74} (1995) 2885.
\bibitem{Ciu:94} {\sc M. Ciuchini, E. Franco, G.Martinelli, L. Reina
and L. Silvestrini,}
{\sl Phys.~Lett.} {\bf B334} (1994) 137.
\bibitem{MM95}
{\sc M. M\"unz}, PhD Thesis, Technical University Munich (1995).
\bibitem{ALI95}
{\sc A. Ali,} {\bf DESY 95-157, hep-ph/9508335}
\bibitem{HWS:87}
{\sc {W. S. Hou, R. I. Willey and A. Soni}},
{\sl Phys.~Rev.~Lett.} {\bf 58} (1987) 1608.
\bibitem{GSW:89}
{\sc B. Grinstein, M. J. Savage and M. B. Wise,}
{\sl Nucl.~Phys.} {\bf B319} (1989) 271.
\bibitem{GDSN:89}
{\sc R. Grigjanis, P. J. O'Donnell, M. Sutherland and H. Navelet,}
{\sl Phys.~Lett.} {\bf B223} (1989) 239.
\bibitem{CRV:91}
{\sc G. Cella, G. Ricciardi and A. Vicer{\'e},}
{\sl Phys.~Lett.} {\bf B258} (1991) 212.
\bibitem{FALK}
{\sc A. Falk, M. Luke and M.J. Savage,}
{\sl Phys. Rev.} {\bf D 49} (1994) 3367.
\bibitem{BSEE}
{\sc A. Ali, G.F. Giudice and T. Mannel,} CERN-TH 7346/94,
hep-ph/9408213;
{\sc C. Greub, A. Ioannissian and D. Wyler},
{\sl Phys.~Lett.} {\bf B346} (1995) 149;
{\sc G. Burdman,} Fermilab-Pub-95/113-T, hep-ph/9505352.
\bibitem{MG}
{\sc M. Gronau,} these proceedings.
\bibitem{NQ}
{\sc Y. Nir and H.R. Quinn} in " B Decays ", ed S. Stone
 (World Scientific, 1994),
p. 520; {\sc I. Dunietz,} ibid p.550 and refs. therein.
\bibitem{DUN}
{\sc M. Gronau,} {\sl Phys. Rev. Lett.} {\bf 63} (1989) 1451,
{\sl Phys. Lett.} {\bf B 233} (1989) 479;
{\sc I. Dunietz and J.L Rosner,}{\sl Phys. Rev.}
 {\bf D 34} (1986) 1404;
{\sc R. Aleksan, I. Dunietz, B. Kayser and F. Le Diberder,}
{\sl Nucl. Phys.} {\bf B 361} (1991) 141.
\bibitem{Wyler}
{\sc M. Gronau and D. Wyler,} {\sl Phys. Lett.} {\bf B 265} (1991) 172.
\bibitem{CPASYM}
{\sc M. Gronau and D. London,} {\sl Phys. Rev. Lett.}
 {\bf 65} (1990) 3381,
{\sl Phys. Lett.} {\bf B 253} (1991) 483;
{\sc Y. Nir and H. Quinn,} {\sl Phys. Rev.} {\bf D 42} (1990) 1473,
{\sl Phys. Rev. Lett.} {\bf 67} (1991) 541;
\bibitem{grl}
{\sc M. Gronau, J.L. Rosner and D. London,} {\sl Phys. Rev. Lett.}
{\bf 73} (1994) 21.
\bibitem{ghlr}
{\sc M. Gronau, O.F. Hern\'andez, D. London and J.L. Rosner},
{\sl Phys. Rev.} {\bf D50} (1994) 4529.
\bibitem{hlgr}
{\sc O.F. Hern\'andez, D. London, M. Gronau and J.L. Rosner,}
{\sl Phys. Lett.} {\bf B333} (1994) 500.
\bibitem{BSANDA}
{\sc I.I.Y. Bigi and A.I. Sanda,}
{\sl Nucl. Phys.} {\bf B 193} (1981) 85.
\bibitem{PAPII}
{\sc A.J. Buras and R. Fleischer,}
{\bf TUM-T31-96/95, hep-ph/9507460} (1995).
\bibitem{KP}
{\sc G. Kramer and W.F. Palmer},
{\bf DESY 95-131, hep-ph/9507329} (1995).
\bibitem{dhewp2}
{\sc N.G. Deshpande and X.-G. He,} {\sl Phys. Rev. Lett.}
{\bf 74} (1995) 26.
\bibitem{ghlrewp}
{\sc M. Gronau, O.F. Hern\'andez, D. London and J.L. Rosner,}
{\bf TECHNION-PH-95-11}, {\bf hep-ph/9504327} (1995).
\bibitem{PAPI}
{\sc A.J. Buras and R. Fleischer,}
{\bf TUM-T31-95/95, hep-ph/9507303} (1995).
\bibitem{PAPIII}
{\sc R. Fleischer,}
{\bf TTP95-32, hep-ph/9509204.} (1995).
\bibitem{DUN2}
{\sc I. Dunietz}, {\sl Phys. Lett.} {\bf B270} (1991) 75.
\bibitem{adk}
{\sc R. Aleksan, I. Dunietz and B. Kayser,}
 {\sl Z.Phys.} {\bf C54} (1992) 653.
\bibitem{DUNS}
{\sc I. Dunietz,} Snowmass B-Physics 1993, p.83.
\bibitem{bf}
{\sc A.J. Buras and R. Fleischer,}
{\sl Phys. Lett.} {\bf B341} (1995) 379.
\bibitem{ghlrsu3}
{\sc M. Gronau, O.F. Hern\'andez, D. London and J.L. Rosner,}
{\bf TECHNION-PH-95-10}, {\bf hep-ph/9504326} (1995).
\bibitem{rfewp1}
{\sc R. Fleischer}, {\sl Z.Phys.} {\bf C62} (1994) 81.
\bibitem{rfewp2}
{\sc R. Fleischer,}
{\sl Phys. Lett.} {\bf B321} (1994) 259.
\bibitem{rfewp3}
{\sc R. Fleischer}, {\sl Phys. Lett.} {\bf B332} (1994) 419.
\bibitem{dhewp}
{\sc N.G. Deshpande and X.-G. He,},
{\sl Phys. Lett.} {\bf B 336} (1994) 471.
\bibitem{dht}
{\sc N.G. Deshpande, X.-G. He and J. Trampetic,}
{\sl Phys. Lett.} {\bf B345} (1995) 547.
\bibitem{dy}
{\sc D. Du and M. Yang,}
{\bf BIHEP-TH-95-8}, {\bf hep-ph/9503278} (1995).
\bibitem{dhgam}
{\sc N.G. Deshpande and X.-G. He,}
{\bf OITS-576}, {\bf hep-ph/9505369} (1995).
\bibitem{PAPF}
{\sc R. Fleischer}, {\sl Phys. Lett.} {\bf B 341} (1994) 205.
\bibitem{WO1}
{\sc L. Wolfenstein}, {\sl Phys. Rev. Lett.} {\bf 13} (1964) 562.
\bibitem{WIN}
{\sc B. Winstein}, {\sl Phys. Rev. Lett.} {\bf 68} (1992) 1271.
\bibitem{AB94}
{\sc A.J. Buras,} {\sl Phys. Lett.} {\bf B 333} (1994) 476.
\bibitem{AB94A}
{\sc A.J. Buras,} {\sl Acta Physica Polonica}, {\bf 26} (1995) 755.
\bibitem{BB4}
{\sc G. Buchalla and A.J. Buras,} {\sl Phys. Lett.} {\bf B 333} (1994) 221.
\end{thebibliography}
\end{document}